\def\k{\kappa}

\def\m{\mu}
\def\n{\nu}

\def\ha{{1\over 2}}
\newcommand{\kl}[3]{\mbox{$\rm #1$}^{\mu\nu , \alpha\beta}_{#2}(#3)}

\def\be{\begin{equation}}
\def\ee{\end{equation}}
\def\te{\end{equation}}
\def\bea{\begin{eqnarray}}

\def\eea{\end{eqnarray}}
\def\tea{\end{eqnarray}}

\def\ha{{1\over 2}}

\documentstyle[11pt]{article}

\makeatletter                    
\@addtoreset{equation}{section}  
\makeatother                     

\begin{document}
\title{Stochastic Gravity: A Primer with Applications}
\author{B. L. Hu
    \thanks{Electronic address: {\tt hub@physics.umd.edu}}\\
  {\small Department of Physics, University of Maryland,
             College Park, Maryland 20742-4111, U.S.A.}\\
E. Verdaguer
    \thanks{Electronic address: {\tt verdague@ffn.ub.es,
              enric@physics.umd.edu}}\\
 {\small Departament de Fisica Fonamental and C.E.R. en Astrofisica
Fisica de Particules i Cosmologia,}\\
           {\small  Universitat de Barcelona,
             Av.~Diagonal 647, 08028 Barcelona, Spain}\\}
\date{({\scriptsize umdpp 03-21, Nov. 26, 2002. Intended as a review in {\it Classical and Quantum Gravity}})}
\maketitle
\begin{abstract}
Stochastic semiclassical gravity of the 90's is a theory naturally
evolved  from semiclassical gravity of the 70's and 80's. It
improves on the semiclassical Einstein equation with source given
by the expectation value of the stress-energy tensor of quantum
matter fields in curved spacetimes by incorporating an additional
source due to their fluctuations. In  stochastic semiclassical
gravity the main object of interest is the noise kernel, the
vacuum expectation value of the (operator-valued) stress-energy
bi-tensor, and the centerpiece is the (stochastic)
Einstein-Langevin equation. We describe this new theory  via two
approaches: the axiomatic and the functional. The axiomatic
approach is useful to see the structure of the theory from the
framework of semiclassical gravity, showing the link from the
mean value of the energy momentum tensor to their correlation
functions. The functional approach uses the Feynman-Vernon
influence functional and the Schwinger-Keldysh close-time-path
effective action methods which are convenient for computations.
It also brings out the open systems concepts and the statistical
and stochastic contents of the theory such as dissipation,
fluctuations, noise and decoherence. We then describe the
applications of stochastic gravity to the backreaction problems
in cosmology and black hole physics. In the first problem we
study the backreaction of conformally coupled quantum fields in a
weakly inhomogeneous cosmology. In the second problem we study the
backreaction of a thermal field in the gravitational background
of a quasi-static black hole (enclosed in a box) and its
fluctuations. These examples serve to illustrate closely the
ideas and techniques presented in the first part. This article is
intended as a first introduction providing readers with some basic
ideas and working knowledge. Thus we place more emphasis here on
pedagogy than completeness. [Further discussions of ideas, issues
and on-going research topics can be found in
\cite{stogra,HVErice,HVLivRev} respectively.]
\end{abstract}


\newpage

\centerline{\bf THEORY: From Mean Values to Fluctuations}


\section{From Semiclassical to Stochastic Gravity }
\label{sec1}

The first step in the road to stochastic gravity begins with
{\it{quantum field theory in curved spacetimes}} (QFTCST)
~\cite{DeW75,BirDav,Fulling89,Wald94,mostepanenko}, which
describes the behavior of quantum matter fields propagating in a
specified (not dynamically determined by the quantum matter field
as source) background gravitational field. In this theory the
gravitational field is given by the classical spacetime metric
determined from classical sources by the classical Einstein
equations, and the quantum fields propagate as test fields in
such a spacetime. For time dependent spacetime geometry it may
not be possible to define a physically meaningful vacuum state
for the quantum field at all times. Assuming there is an initial
time  the vacuum state at a latter time will differ from that
defined initially because particles are created in the
intervening time. An important process described by QFTCST is
indeed particle creation from the vacuum (and effects of vacuum
fluctuations and polarizations) in the early universe \cite{cpc}
and Hawking radiation in black holes \cite{bhpc,bhpc1}.

The second step in the description of the interaction of gravity
with quantum fields is back-reaction, {\it i.e.}, the effect of
the quantum fields on the spacetime geometry. The source here is
the expectation value of the stress-energy operator for the
matter fields in some quantum state in the spacetime, a classical
observable. However, since this object is quadratic in the field
operators, which are only well defined as distributions on the
spacetime, it involves ill defined quantities. It contains
ultraviolet divergences the removal of which requires a
renormalization procedure. The ultraviolet divergences are
already present in Minkowski spacetime, but in a curved
background the renormalization procedure is more involved since
it needs to preserve general covariance \cite{DeW75,Christensen}.
The final expectation value of the stress-energy operator using a
reasonable regularization technique is essentially unique, modulo
some terms which depend on the spacetime curvature and which are
independent of the quantum state. This uniqueness was proved by
Wald \cite{Wald77,Wald78} who investigated the criteria that a
physically meaningful expectation value of the  stress-energy
tensor ought to satisfy.

The theory obtained from a self-consistent solution of the
geometry of the spacetime and the quantum field is known as {\it
semiclassical gravity}. Incorporating the backreaction of the
quantum matter field on the spacetime is thus the central task in
semiclassical gravity. One assumes a general class of spacetime
where the quantum fields live in and act on, and seek a solution
which satisfies simultaneously the Einstein equation for the
spacetime and the field equations for the quantum fields. The
Einstein equation which has the expectation value of the
stress-energy operator of the quantum matter field as the source
is known as the {\it semiclassical Einstein equation} (SEE).
Semiclassical gravity was first investigated in cosmological
backreaction problems \cite{cpcbkr,Hartle1}, an example is the
damping of anisotropy in Bianchi universes by the backreaction of
vacuum particle creation. Using the effect of quantum field
processes such as particle creation to explain why the universe
is so isotropic at the present was investigated in the context of
chaotic cosmology \cite{mixmaster} in the late seventies prior to
the inflationary cosmology proposal of the eighties \cite{infcos},
which assumes the vacuum expectation value of a Higg's field as
the source, another, perhaps more well-known, example of
semiclassical gravity.

\subsection{The importance of quantum fluctuations}

For a free quantum field semiclassical gravity is robust in the
sense that it is consistent and fairly well understood. The
theory is in some sense unique, note that the only reasonable
c-number stress-energy tensor that one may construct
\cite{Wald77,Wald78} with the stress-energy operator is a
renormalized expectation value. However the scope and limitations
of the theory are not so well understood. It is expected that the
semiclassical theory would  break down at the Planck scale. One
can conceivably assume that it would also break down when the
fluctuations of the stress-energy operator are large
\cite{Ford82,KuoFor}. Calculations of the fluctuations of the
energy density for Minkowski, Casimir and hot flat spaces as well
as Einstein and de Sitter universes are available
\cite{KuoFor,PH97,HP0,PH0,PH1,PH2,PH3,MV1,MV2,RV99,RV02a,OsbSho,CEZ}.
It is less clear, however, how to quantify what a large
fluctuation is, and different criteria have been proposed
\cite{KuoFor,ForSCG,ForWu,HP0,PH0,AMM02}. The issue of the
validity of the semiclassical gravity viewed in the light of
quantum fluctuations is summarized in our Erice lectures
\cite{HVErice}. One can see the essence of the problem by the
following example inspired by Ford \cite{Ford82}.

Let us assume a quantum state formed by an isolated system which
consists of a superposition with equal amplitude of  one
configuration of mass $M$ with the center of mass at $X_1$ and
another configuration of the same mass with the center of mass at
$X_2$. The semiclassical theory as described by the semiclassical
Einstein equation predicts that the center of mass of the
gravitational field of the system is centered at $(X_1+X_2)/2$.
However, one would expect that if we send a succession of test
particles to probe the gravitational field of the above system
half of the time they would react to a gravitational field of
mass $M$ centered at $X_1$ and half of the time to the field
centered at $X_2$. The two predictions are clearly different,
note that the fluctuation in the position of the center of masses
is of the order of $(X_1-X_2)^2$. Although this example raises
the issue of how to place the importance of fluctuations to the
mean, a word of caution should be added to the effect that it
should not be taken too literally. In fact, if the previous
masses are macroscopic the quantum system decoheres very quickly
\cite{Zurek91} and instead of being described by a pure quantum
state it is described by a density matrix which diagonalizes in a
certain pointer basis. For observables associated to such a
pointer basis the density matrix description is equivalent to
that provided by a statistical ensemble. The results will differ,
in any case, from the semiclassical prediction.

In other words, one would expect that a stochastic source that
describes the quantum fluctuations should enter into the
semiclassical equations. A significant step in this direction was
made in Ref.~\cite{Physica} where it was proposed to view the
back-reaction problem in the framework of an open quantum system:
the quantum fields seen as the ``environment" and the
gravitational field as the ``system". Following this proposal a
systematic study of the connection between semiclassical gravity
and open quantum systems resulted in the development of a new
conceptual and technical framework where (semiclassical)
Einstein-Langevin equations were derived
\cite{CH94,HM3,HuSin,CamVer96,LomMaz97} \footnote{The word
semiclassical put in parentheses here refers to the fact that the
noise source in the Langevin equation arises from the quantum
field, while the background spacetime is classical. We will not
carry this word in connection with the ELE or stochastic gravity,
since there is no confusion that the source which contributes to
the stochastic features of this theory comes from quantum
fields.} The key technical factor to most of these results was
the use of the influence functional method of Feynman and Vernon
\cite{FeyVer} when only the coarse-grained effect of the
environment on the system is of interest.

However,  although  Einstein-Langevin equations were derived for
several models, the results were somewhat formal and some concern
could be raised on the physical reality of the solutions of the
stochastic equations for the gravitational field. This is related
to the issue of the environment induced quantum to classical
transition. In the language of the consistent histories
formulation of quantum mechanics \cite{conhis} for the existence
of a semiclassical regime for the dynamics of the system one
needs two requirements: The first is decoherence, which
guarantees that probabilities can be consistently assigned to
histories describing the evolution of the system, and the second
is that these probabilities should peak near histories which
correspond to solutions of classical equations of motion. The
effect of the environment is crucial, on the one hand, to provide
decoherence and, on the other hand, to produce both dissipation
and noise to the system through back-reaction, thus inducing a
semiclassical stochastic dynamics on the system. As shown by
different authors \cite{gell-mann-hartle,envdec} (indeed over a
long history predating the current revival of decoherence)
stochastic semiclassical equations are obtained in an open
quantum system after a coarse graining of the environmental
degrees of freedom and a further coarse graining in the system
variables. It is expected but has not yet been shown that this
mechanism could also work for decoherence and classicalization of
the metric field, since a quantum description of the
gravitational field is lacking. Thus far, the analogy could be
made formally \cite{MV3} or under certain assumptions (such as
adopting the Born-Oppenheimer approximation in quantum cosmology
\cite{PazSin}).

Later an axiomatic approach to the Einstein-Langevin equation
without invoking the open system analogy was suggested based on
the formulation of self-consistent dynamical equation for a
perturbative extension to semiclassical gravity able to account
for the lowest order stress-energy fluctuations of matter fields
\cite{MV0}. It was then shown that the same equation could be
derived, in this general case, from the influence functional of
Feynman and Vernon \cite{MV1}. The field equation is deduced via
an effective action which is computed assuming that the
gravitational field is a c-number. It is interesting to note that
the Einstein-Langevin equation can also be understood as a useful
intermediary tool to compute symmetrized two-point correlations
of the quantum metric perturbations on the semiclassical
background, independent of a suitable classicalization mechanism
\cite{RV02b}. The important new  element in the derivation of the
Einstein-Langevin equation, and of the stochastic gravity theory,
is the physical observable that measures the stress-energy
fluctuations, namely, the expectation value of the symmetrized
bi-tensor constructed with the stress-energy tensor operator: the
{\it noise kernel}.

\subsection{An illustrative model}

Before embarking on the formulation of stochastic gravity let us
illustrate the theory with a simple toy model which  minimalizes
the technical complications. The model will be useful to clarify
the role of the noise kernel and illustrate the relationship
between the semiclassical, stochastic and quantum descriptions.
Let us assume that the gravitational equations are described by a
linear field $h(x)$ whose source is a massless scalar field
$\phi(x)$ which satisfies the Klein-Gordon equation in flat
spacetime $\Box \phi(x)=0$. The field stress-energy tensor is
quadratic in the field, and independent of $h(x)$. The classical
gravitational field equations will be given by \footnote{In this
article we use the $(-,+,+,+)$ sign conventions of Refs.
\cite{MTW,Wald84}, and units in which
$c=\hbar=1$.}\begin{equation} \Box h(x)=\kappa T(x), \label{2.12a}
\end{equation}
where $T(x)$ is the (scalar) trace of the stress-energy tensor,
$T(x)=\partial_a\phi(x)\partial^a\phi(x)$ and $\kappa \equiv 16
\pi G$, where G is Newton's constant. Note that this is not a
self-consistent theory since $\phi(x)$ does not react to the
gravitational field $h(x)$. We should also emphasize that this
model is not the standard linearized theory of gravity in which
$T$ is also linear in $h(x)$. It captures, however, some of the
key features of linearized gravity.

In the Heisenberg representation the quantum field $\hat h(x)$ satisfies
\begin{equation}
\Box \hat h(x)=\kappa\hat T(x).
\label{2.12}
\end{equation}
Since $\hat T(x)$ is quadratic in the field operator
$\hat\phi (x)$ some regularization procedure has to be assumed in order
for (\ref{2.12}) to make sense. Since we work in flat spacetime we may
simply use a normal ordering prescription to regularize the operator $\hat
T(x)$. The solutions of this equation, i.e. the field operator at the
point $x$, $\hat h(x)$,  may be written in terms of the retarded
propagator $G(x,y)$ as, \begin{equation} \hat h(x)=\hat
h^0(x)+{1\over\kappa}\int dx' G(x,x^\prime)\hat T(x^\prime),
\label{2.13}
\end{equation}
where $\hat h^0(x)$ is the free field which carries information on the
inital conditions and the state of the field. {}From this solution we may
compute, for instance, the symmetric two point quantum
correlation function (the anticommutator)
\begin{equation}
{1\over2}\langle \{\hat h(x),\hat h(y)\}\rangle = {1\over2}
\langle \{\hat h^0(x),\hat h^0(y)\}\rangle + {1\over 2\kappa^2}
\int\int dx'dy'G(x,x^\prime)G(y,y^\prime) \langle\{\hat
T(x^\prime),\hat T(y^\prime)\}\rangle, \label{2.14}
\end{equation}
where
the expectation value is taken with respect to the quantum state
in which both fields $\phi(x)$ and $h(x)$ are quantized. (We
assume for the free field, $\langle \hat h^0\rangle=0$.)

We can now consider the semiclassical theory for this problem.
If we assume that $h(x)$ is classical and the matter field is quantum the
semiclassical theory may just be described by substituting into the
classical equation (\ref{2.12a}) the stress-energy trace by the
expectation value of the stress-energy trace operator $\langle \hat
T(x)\rangle$, in some quantum state of the field $\hat \phi(x)$.  Since in
our model $\hat T(x)$ is independent of $h(x)$ we may simply renormalize
its expectation value using normal ordering, then for the vacuum state of
the field $\hat\phi(x)$, we would simply have $\langle\hat T(x)\rangle_0=0
$. The semiclassical theory thus reduces to
\begin{equation}
\Box  h(x)=\kappa \langle \hat T(x)\rangle.
\label{2.15a}
\end{equation}
The two point function $h(x)h(y)$ that one may derive from this equation
depends on the two point function $\langle \hat T(x)\rangle \langle \hat
T(y)\rangle $ and clearly cannot reproduce the quantum result
(\ref{2.14}) which depends on the expectation value of
two point operator $\langle\{\hat T(x),\hat
T(y)\}\rangle$. That is, the semiclassical theory entirely misses the
fluctuations of the stress-energy operator $\hat T(x)$.

Let us now see how we can extend the semiclassical theory in order
to account for such fluctuations. The first step is to characterize
these fluctuations. For this, we introduce the noise kernel as the physical
observable that measures the fluctuations of the stress-energy operator
$\hat T$. Define
\begin{equation}
N(x,y)= \frac{1}{2}\langle\{\hat
t(x),\hat t(y)\}\rangle
\label{2.14a}
\end{equation}
where $\hat t(x)=\hat T(x)-\langle\hat T(x)\rangle$. The
bi-scalar $N(x,y)$ is real and positive-semidefinite, as a
consequence of $\hat t$ being self-adjoint. A simple proof can be
given as follows. Let $|\psi\rangle$ be a given quantum state and
let $\hat Q$ be a self-adjoint operator, $\hat Q^\dagger=\hat Q$,
then one can write $\langle\psi|\hat Q\hat Q|\psi\rangle=
\langle\psi|\hat Q^\dagger Q|\psi\rangle= | \hat
Q|\psi\rangle|^2\geq 0$. Now let $\hat t(x)$ be a self-adjoint
operator, then if we define $\hat Q=\int dx f(x) \hat t(x)$ for
an arbitrary well behaved function $f(x)$, the previous inequality
can be written as $\int dx dy f(x)\langle\psi|\hat t(x)\hat
t(y)|\psi\rangle f(y)\geq 0$, which is the condition for the
noise kernel to be positive semi-definite. Note that when
considering the inverse kernel $N^{-1}(x,y)$, it is implicitly
assumed that one is working in the subspace obtained from the
eigenvectors which have strictly positive eigenvalues when the
noise kernel is diagonalized.

By the positive semi-definite property of the noise kernel $N(x,y)$ it is
possible to introduce a Gaussian stochastic field as follows:
\begin{equation}
\langle\xi(x)\rangle_s=0,\quad \langle\xi(x)\xi(y)\rangle_s=N(x,y).
\label{2.14b}
\end{equation}
where the subscript $s$ means a statistical average.
These equations entirely define the stochastic process $\xi(x)$ since we
have assumed that it is Gaussian. Of course, higher correlations could also
be introduced but we just try to capture the fluctuations to lowest order.

The extension of the semiclassical equation may be simply performed by
adding to the right-hand side of the semiclassical equation (\ref{2.15a})
this stochastic source $\xi(x)$ which accounts for the fluctuations of
$\hat T$ as follows,
\begin{equation}
\Box  h(x)=\kappa\left( \langle \hat T(x)\rangle+\xi(x)\right).
\label{2.15}
\end{equation}
This equation is in the form of a Langevin equation: the field
$h(x)$ is classical but stochastic and the observables we may
obtain from it are correlation functions for $h(x)$. In fact, the
solution of this equation may be written in terms of the retarded
propagator as, \begin{equation} h(x)=h^0(x)+{1\over\kappa}\int
dx^\prime G(x,x^\prime)\left(\langle\hat T(x^\prime)\rangle
+\xi(x^\prime)\right) , \label{2.16}
\end{equation}
from where the two point correlation function for the classical field
$h(x)$, after using the definition of $\xi(x)$ and that $\langle
h^0(x)\rangle_s=0$, is given by
\begin{equation}
\langle  h(x) h(y)\rangle_s = \langle  h^0(x) h^0(y)\rangle_s
+{1\over2\kappa^2} \int\int dx^\prime dy^\prime
G(x,x^\prime)G(y,y^\prime) \langle\{\hat T(x^\prime),\hat
T(y^\prime)\}\rangle. \label{2.17}
\end{equation}
Note that in writing $\left<\dots\right>_s$ here we are assuming
a double stochastic average, one is related to the stochastic
field $\xi(x)$ and the other is related to the free field
$h^0(x)$ which is assumed also to be stochastic with a
distribution function to be specified.

Comparing (\ref{2.14}) with (\ref{2.17}) we see that the
respective second term on the right-hand side are identical
provided the expectation values are computed in the same quantum
state for the field $\hat \phi(x)$ (recall that we have assumed
$T(x)$ does not depend on $h(x)$). The fact that the field $h(x)$
is also quantized in (\ref{2.14}) does not change the previous
statement. The nature of the first term on the right-hand sides
of equations (\ref{2.14}) and (\ref{2.17}) is different: in the
first case it is the two point quantum expectation value of the
free quantum field $\hat h^0$ whereas in the second case it is the
stochastic average of the two point classical homogeneous field
$h^0$, which depends on the initial conditions. Now we can still
make these terms equal to each other if we assume for the
homogeneous field $h^0$ a Gaussian distribution of initial
conditions such that
\begin{equation}
\langle  h^0(x)
h^0(y)\rangle_s=  \frac{1}{2}\langle\{\hat h^0(x),\hat
h^0(y)\}\rangle.
\label{2.17a}
\end{equation}
This Gaussian stochastic field $h^0(x)$ can always be defined due
to the positivity of the anti-commutator. Thus, under this
assumption on the initial conditions for the field $h(x)$ the two
point correlation function of (\ref{2.17}) equals the quantum
expectation value of (\ref{2.14}) exactly. An interesting feature
of the stochastic description is that the quantum anticommutator
of (\ref{2.14}) can be written as the right-hand side of equation
(\ref{2.17}) where the first term contains all the information on
initial conditions for the stochastic field $h(x)$ and the second
term codifies all the information on the quantum correlations of
the source. This separation is also seen in the description of
some quantum Brownian motion models which are typically used as
paradigms of open quantum systems
\cite{CRVopensys,CRVnoisetunnel}.

It is interesting to note that in the standard linearized theory
of gravity $T(x)$ depends also on $h(x)$, both explicitly and also
impicitly through the coupling of $\phi(x)$ with $h(x)$. The
equations are not so simple but it is still true that the
corresponding Langevin equation leads to the correct symmetrized
two point quantum correlations for the metric perturbations
\cite{MV2,RV02b}. Thus in a linear theory as in the model just
described one may just use the statistical description given by
(\ref{2.15}) to compute the symmetric quantum two point function
of equation (\ref{2.13}). This does not mean that we can recover
all quantum correlation functions with the stochastic
description, see Ref. \cite{CRVopensys} for a general discussion
about this point. Note that, for instance, the commutator of the
classical stochastic field $h(x)$ is obviously zero, but the
commutator of the quantum field $\hat h(x)$ is not zero for
timelike separated points; this is the prize we pay for the
introduction of the classical field $\xi(x)$ to describe the
quantum fluctuations. Furthermore, the statistical description is
not able to account for the graviton-graviton effects which go
beyond the linear approximation in $\hat h(x)$.

\section{The Einstein-Langevin equation: Axiomatic approach}
\label{sec2}

In this section we introduce {\it stochastic semiclassical
gravity} in an axiomatic way, following closely the previous toy
model. It is introduced as an extension of semiclassical gravity
motivated by the search of self-consistent equations which
describe the back-reaction of the quantum stress-energy
fluctuations on the gravitational field \cite{MV0}.

\subsection{Semiclassical gravity}

Semiclassical gravity describes the interaction of a classical
gravitational field with quantum matter fields. This theory can
be formally derived as the leading $1/N$ approximation of quantum
gravity interacting with $N$ independent and identical free
quantum fields \cite{horowitz-wald,Hartle-Horowitz81,Tomboulis77}
which interact with gravity only.  By keeping the value of $NG$
finite, where $G$ is Newton's gravitational constant, one arrives
at a theory in which formally the gravitational field can be
treated as a c-number field (i.e. quantized at tree level) while
matter fields are fully quantized.

The semiclassical theory may be summarized as follows.
Let $({\cal M},g_{ab})$ be a globally hyperbolic four-dimensional
spacetime manifold ${\cal M}$ with metric $g_{ab}$ and consider a real
scalar quantum field $\phi$ of mass $m$ propagating on that manifold;
we just assume a scalar field for
simplicity.
The classical action $S_m$ for this matter field is given by
the functional
\begin{equation}
S_m[g,\phi]=-{1\over2}\int d^4x\sqrt{-g}\left[g^{ab}
\nabla_a\phi\nabla_b\phi+\left(m^2+\xi R\right)\phi^2\right],
\label{2.1}
\end{equation}
where $\nabla_a$ is the covariant derivative associated to the
metric $g_{ab}$, $\xi$ is a  coupling parameter between the field
and the scalar curvature of the underlying spacetime $R$, and
$g={\rm det} g_{ab}$.

The field may be quantized in the manifold using the standard
canonical quantization formalism \cite{BirDav,Fulling89,Wald94}.
The field operator in the Heisenberg representation $\hat\phi$ is
an operator-valued distribution solution of the Klein-Gordon
equation, the field equation derived from  Eq. (\ref{2.1}),
\begin{equation}
(\Box-m^2 -\xi R)\hat\phi=0.
\label{2.2}
\end{equation}
We may write the field operator as $\hat\phi[g;x)$
to indicate that it is a functional of the metric $g_{ab}$ and a function
of the spacetime point $x$. This notation will be used also for other
operators and tensors.

The classical stress-energy tensor is obtained by functional derivation
of this action in the usual way
$T^{ab}(x)=(2/\sqrt{-g})\delta S_m/\delta g_{ab}$, leading to
\begin{eqnarray}
T^{ab}[g,\phi]&=&\nabla^a\phi\nabla^b\phi-{1\over2}g^{ab}
\left(\nabla^c\phi\nabla_c\phi+m^2\phi^2\right)
\nonumber \\
&&+\xi\left(g^{ab}\Box-\nabla^a\nabla^b+G^{ab}
\right)\phi^2,
\label{2.3}
\end{eqnarray}
where $\Box=\nabla_a\nabla^a$ and $G_{ab}$ is the Einstein tensor.
With the notation $T^{ab}[g,\phi]$ we explicitly
indicate that the stress-energy tensor is
a functional of the metric $g_{ab}$  and the field $\phi$.

The next step is to define a stress-energy tensor operator $\hat
T^{ab}[g;x)$. Naively one would replace the classical field
$\phi[g;x)$ in the above functional by the quantum operator
$\hat\phi[g;x)$, but this procedure involves taking the product
of two distributions at the same spacetime point. This is
ill-defined and we need a regularization procedure. There are
several regularization methods which one may use, one is the
point-splitting or point-separation regularization method
\cite{Christensen} in which one introduces a point $y$ in a
neighborhood of the point $x$ and then uses as the regulator the
vector tangent at the point $x$ of the geodesic joining $x$ and
$y$; this method is discussed for instance in Ref.
\cite{PH1,PH2,PH3}. Another well known method is dimensional
regularization in which one works in arbitrary $n$ dimensions,
where $n$ is not necessarily an integer, and then uses as the
regulator the parameter $\epsilon=n-4$; this method is implicitly
used in this section. The regularized stress-energy operator
using the Weyl ordering prescription, {\it i.e.} symmetrical
ordering, can be written as \be \hat{T}^{ab}[g] = {1\over 2} \{
     \nabla^{a}\hat{\phi}[g]\, , \,
     \nabla^{b}\hat{\phi}[g] \}
     + {\cal D}^{ab}[g]\, \hat{\phi}^2[g],
\label{regul s-t 2} \ee where ${\cal D}^{ab}[g]$ is the
differential operator: $ {\cal D}^{ab} \equiv
\left(\xi-1/4\right) g^{ab} \Box+ \xi \left( R^{ab}- \nabla^{a}
\nabla^{b}\right). $ Note that if dimensional regularization is
used, the field operator $\hat \phi[g;x)$ propagates in a
$n$-dimensional spacetime. Once the regularization prescription
has been introduced a regularized and renormalized stress-energy
operator $\hat T^R_{ab}[g;x)$ may be defined which differs from
the regularized $\hat T_{ab}[g;x)$ by the identity operator times
some tensor counterterms, which depend on the regulator and are
local functionals of the metric, see Ref.~\cite{MV1} for details.
The field states can be chosen in such a way that for any pair of
physically acceptable states, {\it i.e.}, Hadamard states in the
sense of Ref.~\cite{Wald94}, $|\psi\rangle$, and
$|\varphi\rangle$ the matrix element
$\langle\psi|T^R_{ab}|\varphi\rangle$, defined as the limit when
the regulator takes the physical value, is finite and satisfies
Wald's axioms \cite{Fulling89,Wald77}. These counterterms can be
extracted from the singular part of a Schwinger-DeWitt series
\cite{Fulling89,Christensen,Bunch79}. The choice of these
counterterms is not unique but this ambiguity can be absorbed
into the renormalized coupling constants which appear in the
equations of motion for the gravitational field.

The {\it semiclassical Einstein equation} for the metric $g_{ab}$
can then be written as
\begin{equation}
G_{ab}[g]+\Lambda g_{ab}
-2(\alpha A_{ab}+\beta B_{ab})[g]=
8\pi G \langle \hat T_{ab}^R[g]\rangle ,
\label{2.5}
\end{equation}
where $\langle \hat T_{ab}^R[g]\rangle $ is the expectation value
of the operator $\hat T_{ab}^R[g,x)$ after the regulator takes
the physical value in some physically acceptable state of the field on
$({\cal M},g_{ab})$. Note that both the stress tensor and the
quantum state are functionals of the metric, hence the notation.
The parameters $G$, $\Lambda$, $\alpha$ and $\beta$ are the
renormalized coupling constants, respectively, the gravitational
constant, the cosmological constant and two dimensionless  coupling
constants which are zero in the classical Einstein equation.
These constants must be understood as the result of  ``dressing''
the bare constants which appear in the classical action
before renormalization. The values of these constants must be
determined by experiment.
The left hand side of Eq. (\ref{2.5}) may be derived
from the gravitational action
\begin{equation}
S_g[g]= {1\over 8\pi G}\int d^4 x \sqrt{-g}\left[ {1\over 2}
R-\Lambda +\alpha C_{abcd}C^{abcd}
+\beta R^2\right],
\label{2.6}
\end{equation}
where $C_{abcd}$ is the Weyl tensor. The tensors
$A_{ab}$ and $B_{ab}$ come from the functional
derivatives with respect to the metric of the terms quadratic
in the curvature in Eq. (\ref{2.6}), they are explicitly given by
\begin{eqnarray}
A^{ab}&=&\frac{1}{\sqrt{-g}}\frac{\delta}{\delta g_{ab}}
\int d^4 \sqrt{-g} C_{cdef}C^{cdef}
\nonumber\\
&=&{1\over2}g^{ab}C_{cdef}
C^{cdef}-2R^{acde}
R^{b}_{\ cde}+4R^{ac}R_c^{\ b}
-{2\over3}RR^{ab}
\nonumber\\
&& -2\Box R^{ab}+{2\over3}\nabla^a\nabla^b R+
{1\over3}g^{ab}\Box R,
\label{2.7a}\\
B^{ab}&=&\frac{1}{\sqrt{-g}}\frac{\delta}{\delta g_{ab}}
\int d^4 \sqrt{-g} R^2
\nonumber\\
&=&{1\over2}g^{ab}R^2-2R R^{ab}
+2\nabla^a\nabla^b R-2g^{ab}\Box R,
\label{2.7b}
\end{eqnarray}
where $R_{abcd}$  and
$R_{ab}$ are the Riemann and Ricci tensors, respectively.
These two tensors are, like the Einstein and metric tensors,
symmetric and
divergenceless: $\nabla^a A_{ab}=0=\nabla^a B_{ab}$.
Note that a classical stress-energy tensor can also
be added to the right hand side of Eq.
(\ref{2.5}), but  we omit such a term for simplicity.

A solution of semiclassical gravity consists of a spacetime
(${\cal M},g_{ab}$), a quantum field operator $\hat\phi[g]$
which satisfies the evolution equation (\ref{2.2}), and a physically
acceptable state  $|\psi[g]\rangle $ for this field, such that Eq.
(\ref{2.5}) is satisfied when the expectation value of the renormalized
stress-energy operator is evaluated in this state.

For a free quantum field  this theory is robust in the sense that
it is self-consistent and fairly well understood.
As long as the gravitational field is assumed to be described by a
classical metric, the above semiclassical Einstein
equations seems to be the only plausible dynamical equation
for this metric: the metric couples to matter fields via the
stress-energy tensor and for a given quantum state the only
physically observable c-number  stress-energy tensor that one
can construct is the above renormalized expectation value.
However, lacking a full quantum gravity theory the scope and
limits of the theory are not so well understood. It is assumed
that the semiclassical theory should break down at Planck scales,
which is when simple order of magnitude estimates suggest that
the quantum effects of gravity should not be ignored because the
energy of a quantum fluctuation in
a Planck size region, as determined by the Heisenberg uncertainty
principle, is comparable to the gravitational energy of
that fluctuation.

The theory is expected to break down when the fluctuations of the
stress-energy operator are large \cite{Ford82}. A criterion based
on the ratio of the fluctuations to the mean was proposed by Kuo
and Ford \cite{KuoFor} (see also work via zeta-function methods
\cite{PH97,CEZ}).  This proposal was questioned by Phillips and
Hu \cite{HP0,PH0,PH1} because it does not contain a scale at
which the theory is probed or how accurately the theory can be
resolved. They suggested the use of a smearing scale or
point-separation distance, for integrating over the bi-tensor
quantities, equivalent to a stipulation of the resolution level of
measurements. (See response by Ford \cite{ForSCG,ForWu}). A
different criterion is recently suggested by Anderson et al
\cite{AMM02} based on linear response theory. A partial summary
of this issue can be found in our Erice Lectures \cite{HVErice}.

\subsection{Stochastic gravity}

The purpose of stochastic semiclassical gravity is to extend the
semiclassical theory to account for these fluctuations in a
self-consistent way. A physical observable that describes these
fluctuations to lowest order is the {\it noise kernel} bi-tensor,
which is defined through the two point correlation of the
stress-energy operator as
\begin{equation}
N_{abcd}[g;x,y)={1\over2}\langle\{\hat t_{ab}[g;x),
\hat t_{cd}[g;y)\}\rangle,
\label{2.8}
\end{equation}
where the curly brackets mean anticommutator, and where
\begin{equation}
\hat t_{ab}[g;x)
\equiv \hat T_{ab}[g;x)-\langle \hat T_{ab}[g;x)\rangle.
\label{2.9}
\end{equation}
This bi-tensor is sometimes also written $N_{abc^\prime
d^\prime}[g;x,y)$, or $N_{abc^\prime d^\prime}(x,y)$, to
emphasize that it is a tensor with respect to the first two
indices at the point $x$ and a tensor with respect to the last
two indices at the point $y$, but we shall not follow this
notation here. The noise kernel is defined in terms of the
unrenormalized stress-tensor operator $\hat T_{ab}[g;x)$ on a
given background metric $g_{ab}$, thus a regulator is implicitly
assumed on the right-hand side of Eq. (\ref{2.8}).  However, for
a linear quantum field the above kernel -- the expectation
function of a bi-tensor -- is free of ultraviolet divergences
because the regularized $T_{ab}[g;x)$ differs from the
renormalized $T_{ab}^R[g;x)$ by the identity operator times some
tensor counterterms so that in the substraction (\ref{2.9}) the
counterterms cancel. Consequently the ultraviolet behavior of
$\langle\hat T_{ab}(x)\hat T_{cd}(y)\rangle$ is the same as that
of $\langle\hat T_{ab}(x)\rangle \langle\hat T_{cd}(y)\rangle$,
and $\hat T_{ab}$ can be replaced by the renormalized operator
$\hat T_{ab}^R$ in Eq. (\ref{2.8}); an alternative proof of this
result is given in Ref. \cite{PH1,PH2}. The noise kernel should be
thought of as a distribution function, the limit of coincidence
points has meaning only in the sense of distributions.

The bi-tensor $N_{abcd}[g;x,y)$, or $N_{abcd}(x,y)$ for short, is
real and positive semi-definite, as a consequence of $\hat
T_{ab}^R$ being self-adjoint. A simple proof can be given as in
the toy model example discussed above provided one assumes that
$x$ in that proof carries also tensorial indices.


Once the fluctuations of the stress-energy operator have been
characterized we can  perturbatively extend the semiclassical
theory to account for such fluctuations. Thus we will assume that
the background spacetime metric $g_{ab}$ is a solution of the
semiclassical Einstein Eqs.~(\ref{2.5}) and we will write the new
metric for the extended theory as $g_{ab}+h_{ab}$, where we will
assume that $h_{ab}$ is a perturbation to the background solution.
The renormalized stress-energy operator and the state of the
quantum field may now be denoted by $\hat T_{ab}^R[g+h]$ and
$|\psi[g+h]\rangle$, respectively, and $\langle\hat
T_{ab}^R[g+h]\rangle$ will be the corresponding expectation value.

Let us now introduce a Gaussian stochastic tensor field
$\xi_{ab}[g;x)$ defined by the following correlators:
\begin{equation}
\langle\xi_{ab}[g;x)\rangle_s=0,\ \ \
\langle\xi_{ab}[g;x)\xi_{cd}[g;y)\rangle_s=
N_{abcd}[g;x,y),
\label{2.10}
\end{equation}
where $\langle\dots\rangle_s$ means statistical average. The
symmetry and positive semi-definite property of the noise kernel
guarantees that the stochastic field tensor $\xi_{ab}[g,x)$, or
$\xi_{ab}(x)$ for short, just introduced is well defined. Note
that this stochastic tensor captures only partially the quantum
nature of the fluctuations of the stress-energy operator since it
assumes that cumulants of higher order are zero.

An important property of this stochastic tensor is that it is
covariantly conserved in the background spacetime
$\nabla^a\xi_{ab}[g;x)=0$. In fact, as a consequence of the
conservation of $\hat T_{ab}^R[g]$ one can see that $\nabla_x^a
N_{abcd}(x,y)=0$. Taking the divergence in Eq.~(\ref{2.10}) one
can then show that $\langle\nabla^a\xi_{ab}\rangle_s=0$ and
$\langle\nabla_x^a\xi_{ab}(x) \xi_{cd}(y)\rangle_s=0$ so that
$\nabla^a\xi_{ab}$ is deterministic and represents with certainty
the zero vector field in $\cal{M}$.

For a conformal field, {\it i.e.}, a field whose classical action
is conformally invariant, $\xi_{ab}$ is traceless:
$g^{ab}\xi_{ab}[g;x)=0$; so that, for a conformal matter field
the stochastic source gives no correction to the trace anomaly.
In fact, from the trace anomaly result which states that
$g^{ab}\hat T^R_{ab}[g]$ is in this case a local c-number
functional of $g_{ab}$ times the identity operator, we have that
$g^{ab}(x)N_{abcd}[g;x,y)=0$. It then follows from Eq.
(\ref{2.10}) that $\langle g^{ab}\xi_{ab}\rangle_s=0$ and
$\langle g^{ab}(x)\xi_{ab}(x) \xi_{cd}(y)\rangle_s=0$; an
alternative proof based on the point-separation method is given
in Ref. \cite{PH1,PH2,PH3}, see also Ref. \cite{HVErice}.

All these properties make it quite natural to incorporate into the
Einstein equations the stress-energy fluctuations by using the
stochastic tensor $\xi_{ab}[g;x)$ as the source
of the metric perturbations.
Thus we will write the following equation.
\begin{equation}
G_{ab}[g+h]+
\Lambda (g_{ab}+h_{ab})
-2(\alpha A_{ab}+\beta B_{ab})[g+h]=8\pi G\left(
\langle \hat T_{ab}^R[g+h]\rangle + \xi_{ab}[g]\right).
\label{2.11}
\end{equation}
This equation is in the form of a {\it (semiclassical)
Einstein-Langevin equation}, it is a dynamical equation for the
metric perturbation $h_{ab}$ to linear order. It describes the
back-reaction of the metric to the quantum fluctuations of the
stress-energy tensor of matter fields, and gives a first order
extension to semiclassical gravity as described by the
semiclassical Einstein equation (\ref{2.5}). \footnote{Note that
we refer to the ELE as a first order extension to semiclassical
Einstein equation (SEE) of semiclassical gravity (SCG) and the
lowest level representation of stochastic gravity because in
practice the Feynman-Vernon scheme of representing the noise
kernel by a classical stochastic source is implementable only in
cases where that term in the influence action contains
quadratures of the system variable $h$ [see Eq. (\ref{expansion
2})] via a Gaussian functional integral identity \cite{FeyVer}.
This is possible for linear perturbations but not necessarily so
at higher orders, because of the difficulty in rendering it to a
form amenable to the employ of the Feynman-Vernon identity for
the definition of a classical stochastic source.
However, stochastic gravity has a much broader meaning beyond
these limitations. It refers to the range of theories based on
second and higher order correlation functions. Noise can be
defined in effectively open systems (e.g. correlation noise
\cite{CH00} in the Schwinger-Dyson equation hierarchy) to some
degree but one should not expect the simple Langevin form to
prevail. It is in this sense that we say stochastic gravity is
the intermediate theory between SCG (a mean field theory based on
the expectation values of the energy momentum tensor of quantum
fields) and quantum gravity (the full hierarchy of correlation
functions retaining complete quantum coherence
\cite{stogra,kinQG}}. The renormalization of the operator $\hat
T_{ab}[g+h]$ is carried out exactly as in the previous case, now
in the perturbed metric $g_{ab}+h_{ab}$. Note that the stochastic
source $\xi_{ab}[g;x)$ is not dynamical, it is independent of
$h_{ab}$ since it describes the fluctuations of the stress tensor
on the semiclassical background $g_{ab}$.

An important property of the Einstein-Langevin equation is that it is
gauge invariant under the change of $h_{ab}$ by
$h_{ab}^\prime =h_{ab} +\nabla_a\zeta_b+\nabla_b\zeta_a$, where
$\zeta^a$ is a stochastic vector field on the background manifold ${\cal
M}$. Note that a tensor such as
$R_{ab}[g+h]$, transforms as
$R_{ab}[g+h^\prime]=R_{ab}[g+h]+{\cal L}_\zeta R_{ab}[g]$ to linear order
in the perturbations, where ${\cal L}_\zeta $ is the Lie derivative with
respect to $\zeta^a$. Now, let us write the source tensors in
Eqs.~(\ref{2.11}) and (\ref{2.5}) to the left-hand sides of these
equations. If we substitute  $h$ by $h^\prime$ in this new version of Eq.
(\ref{2.11}), we get the same expression, with $h$ instead of $h^\prime$,
plus the Lie derivative of the combination of tensors which appear on
the left-hand side of the new Eq. (\ref{2.5}). This last combination
vanishes when Eq. (\ref{2.5}) is satisfied, {\it i.e.}, when the
background metric $g_{ab}$ is a solution of semiclassical gravity.

A solution of Eq.~(\ref{2.11}) can be formally written as
$h_{ab}[\xi]$. This solution is characterized by the whole
family of its correlation functions. From the statistical average
of this  equation we have that $g_{ab}+\langle h_{ab}\rangle_s$
must be a solution of the semiclassical Einstein equation linearized
around the background  $g_{ab}$; this solution has been proposed as a test
for the validity of the semiclassical approximation \cite{AMM02}. The
fluctuation of the metric around this average are described by the moments
of the stochastic field $h_{ab}^s[\xi]=h_{ab}[\xi]-\langle
h_{ab}\rangle_s$. Thus the solutions of the Einstein-Langevin equation
will provide the two point metric correlators $\langle h_{ab}^s(x)
h_{cd}^s(y)\rangle_s$.

We see that whereas the semiclassical theory depends on the
expectation value of the point-defined value of the stress-energy
operator, the stochastic theory carries information also on the
two point correlation of the stress-energy operator, as is shown
in the toy model of the previous section, the stochastic theory
may be understood as an intermediate step between the
semiclassical theory based on the mean value, and the full quantum
theory in the sense that it contains extra, yet only partial,
information carried by the n-point functions.

We should also emphasize that, even if the metric fluctuations
appears classical and stochastic, their origin is  quantum not
only because they are induced by the fluctuations of quantum
matter, but also because they are the suitably coarse-grained
variables left over from the quantum gravity fluctuations after
some mechanism for decoherence and classicalization of the metric
field \cite{gell-mann-hartle,hartle,dowker,halliwell,whelan}. One
may, in fact, derive the stochastic semiclassical theory from a
full quantum theory. This was done via the world-line influence
functional method for a moving charged particle in an
electromagnetic field in quantum electrodynamics \cite{JH1}. From
another viewpoint, quite independent of whether a
classicalization mechanism is mandatory or implementable,  the
Einstein-Langevin equation proves to be a useful tool to compute
the symmetrized two point correlations of the quantum metric
perturbations \cite{RV02b}, as illustrated in the simple toy model
described previously.

\section{The Einstein-Langevin equation: Functional approach}
\label{sec3}

The Einstein-Langevin equation (\ref{2.11}) may also be derived by
a method based on functional techniques  \cite{MV1}. In
semiclassical gravity functional methods were used to study the
back-reaction of quantum fields in cosmological models
\cite{Hartle1}. The primary advantage of the effective action
approach is, in addition to the well-known fact that it is easy to
introduce perturbation schemes like loop expansion and nPI
formalisms, that it yields a {\it fully} self-consistent
solution. \footnote{This is because the set of equations of
motion derived from the effective action for {\it both} the
spacetime and the quantum fields are of the same order. Note that
seeking backreaction from the equations of motion level {\it ab
initio} (not derived from the effective action)
straightforwardly, as done in the old ways, i.e., proceeding with
the solution of one {\it after} another would produce results in
staggered orders. To compare these two approaches, equation of
motion versus effective action, see, e.g., the work of Hu and
Parker 1978 versus Hartle and Hu 1979 in \cite{cpcbkr}. See also
comments in Sec. 5.6 on the black hole backreaction problem
comparing the approach by York et al \cite{York} versus that of
Sinha, Raval and Hu \cite{SRH}. The correct way  at the equation
of motion level for stochastic gravity is that illustrated in the
previous section.} The well known in-out effective action method
treated in textbooks, however, led to equations of motion which
were not real because they were tailored to compute transition
elements of quantum operators rather than expectation values. The
correct technique to use for the backreaction problem is the
Schwinger-Keldysh \cite{ctp} closed-time-path (CTP) or `in-in'
effective action. These techniques were adapted to the
gravitational context \cite{ctpcst,CamVer94} and applied to
different problems in cosmology. One could deduce the
semiclassical Einstein equation from the CTP effective action for
the gravitational field (at tree level) with quantum matter
fields.

Furthermore, in this case the CTP functional formalism turns out
to be related
\cite{Su,CH94,CamVer96,LomMaz96,greiner,CamHu,morikawa,MV1} to
the influence functional formalism of Feynman and Vernon
\cite{FeyVer} since the full quantum system may be understood as
consisting of a distinguished subsystem (the ``system'' of
interest) interacting with the remaining degrees of freedom (the
environment). Integration out the environment variables in a CTP
path integral yields the influence functional, from which one can
define an effective action for the dynamics of the system
\cite{CH94,HuSin,HuMat94,greiner}. This approach to semiclassical
gravity is motivated by the observation \cite{Physica} that in
some open quantum systems classicalization and decoherence
\cite{envdec} on the system may be brought about by interaction
with an environment, the environment being in this case the
matter fields and some ``high-momentum'' gravitational modes
\cite{decQC,whelan}. Unfortunately, since the form of a complete
quantum theory of gravity interacting with matter is unknown, we
do not know what these ``high-momentum'' gravitational modes are.
Such a fundamental quantum theory might not even be a field
theory, in which case the metric and scalar fields would not be
fundamental objects \cite{stogra}. Thus, in this case, we cannot
attempt to evaluate the influence action of Feynman and Vernon
starting from the fundamental quantum theory and performing the
path integrations in the environment variables. Instead, we
introduce the influence action for an effective quantum field
theory of gravity and matter \cite{donoghue,SinHu,PazSin}, in
which such ``high-momentum'' gravitational modes are assumed to
have already been ``integrated out.''

Adopting the usual procedure of effective field theories
\cite{weinberg,donoghue,CH97}, one has to take the effective
action for the metric and the scalar field of the most general
local form compatible with general covariance: $S[g,\phi] \equiv
S_g[g]+S_m[g,\phi]+ \cdots ,$ where $S_g[g]$ and $S_m[g,\phi]$
are given by Eqs. (\ref{2.6}) and (\ref{2.1}), respectively, and
the dots stand for terms of order higher than two in the
curvature and in the number of derivatives of the scalar field.
Here, we shall neglect the higher order terms as well as
self-interaction terms for the scalar field. The second order
terms are necessary to renormalize one-loop ultraviolet
divergences of the scalar field stress-energy tensor, as we have
already seen. Since ${\cal M}$ is a globally hyperbolic manifold,
we can foliate it by a family of $t\!=\! {\rm constant}$ Cauchy
hypersurfaces $\Sigma_{t}$, and we will indicate the initial and
final times by $t_i$ and $t_f$, respectively.

The {\it influence functional} corresponding to the action
(\ref{2.1}) describing a scalar field in a spacetime (coupled to
a metric field) may be introduced as a functional of two copies
of the metric, denoted by $g_{ab}^+$ and $g_{ab}^-$, which
coincide at some final time $t=t_f$. Let us assume that, in the
quantum effective theory, the state of the full system (the
scalar and the metric fields) in the Schr\"{o}dinger picture at
the initial time $t\! =\! t_{i}$ can be described by a density
operator which can be written as the tensor product of two
operators on the Hilbert spaces of the metric and of the scalar
field. Let $\rho_i \left[\phi_+(t_i),\phi_-(t_i) \right] $ be the
matrix element of the density operator $\hat{\rho}^{\rm
\scriptscriptstyle S}(t_{i})$ describing the initial state of the
scalar field. The Feynman-Vernon influence functional is defined
as the following path integral over the two copies of the scalar
field:
\begin{equation}
{\cal F}_{\rm IF}[g^\pm] \equiv
\int\! {\cal D}\phi_+\;
{\cal D}\phi_- \;
\rho_i \!\left[\phi_+(t_i),\phi_-(t_i) \right] \,
\delta\!\left[\phi_+(t_f)\!-\!\phi_-(t_f)  \right]\:
e^{i\left(S_m[g^+,\phi_+]-S_m[g^-,\phi_-]\right) }.
\label{path integral}
\end{equation}
Alternatively, the  above double path integral can be rewritten
as a closed time path (CTP) integral, namely, as a single path
integral in a complex time contour with two different time
branches, one going forward in time from $t_i$ to $t_f$, and the
other going backward in time from $t_f$ to $t_i$ (in practice one
usually takes $t_i\to -\infty$). {}From this influence functional,
the {\it influence action} $S_{\rm IF}[g^+,g^-]$, or $S_{\rm
IF}[g^\pm]$ for short,  defined  by \be {\cal F}_{\rm IF}[g^\pm]
\equiv e^{i S_{\rm IF}[g^\pm]}, \label{influence functional} \ee
carries all the information about the environment (the matter
fields) relevant to the system (the gravitational field). Then we
can define the CTP {\it effective action} for the gravitational
field, $S_{\rm eff}[g^\pm]$, as
\begin{equation}
S_{\rm eff}[g^\pm]\equiv S_{g}[g^+]-S_{g}[g^-] +S_{\rm
IF}[g^\pm]. \label{ctpif}
\end{equation}
This is the effective action for the classical gravitational
field in the CTP formalism. However, since the gravitational
field is treated only at the tree level, this is also the
effective classical action from which the classical equations of
motion can be derived.

Expression (\ref{path integral}) contains ultraviolet divergences
and must be regularized. We shall assume that dimensional
regularization can be applied, that is, it makes sense to
dimensionally continue all the quantities that appear in Eq.
(\ref{path integral}).  For this we need to work with the
$n$-dimensional actions corresponding to $S_m$ in (\ref{path
integral}) and $S_g$ in (\ref{2.6}). For example,  the parameters
$G$, $\Lambda$ $\alpha$ and $\beta$ of Eq. (\ref{2.6}) are the
bare parameters $G_B$, $\Lambda_B$, $\alpha_B$ and $\beta_B$, and
in $S_g[g]$, instead of the square of the Weyl tensor in Eq.
(\ref{2.6}),  one must use $(2/3)(R_{abcd}R^{abcd}-
R_{ab}R^{ab})$ which by the Gauss-Bonnet theorem leads to the same
equations of motion as the action (\ref{2.6}) when $n \!=\! 4$.
The form of $S_g$ in $n$ dimensions is suggested by the
Schwinger-DeWitt analysis of the ultraviolet divergences in the
matter stress-energy tensor using dimensional regularization. One
can then write the Feynman-Vernon effective action $S_{\rm
eff}[g^\pm]$ in Eq. (\ref{ctpif}) in a form suitable for
dimensional regularization. Since both $S_m$ and $S_g$ contain
second order derivatives of the metric, one should also add some
boundary terms \cite{Wald84,HuSin}. The effect of these terms is
to cancel out the boundary terms which appear when taking
variations of $S_{\rm eff}[g^\pm]$ keeping the value of
$g^+_{ab}$ and $g^-_{ab}$ fixed at $\Sigma_{t_i}$ and
$\Sigma_{t_f}$. Alternatively, in order to obtain the equations
of motion for the metric in the semiclassical regime, we can work
with the action terms  without boundary terms and neglect all
boundary terms when taking variations with respect to
$g^{\pm}_{ab}$. From now on, all the functional derivatives with
respect to the metric will be understood in this sense.

The semiclassical Einstein equation (\ref{2.5}) can now be derived.
Using the definition of the stress-energy tensor
$T^{ab}(x)=(2/\sqrt{-g})\delta S_m/\delta g_{ab}$
and the definition
of the influence functional, Eqs.
(\ref{path integral}) and (\ref{influence functional}), we see that
\begin{equation}
\langle \hat{T}^{ab}[g;x) \rangle =
\left. {2\over\sqrt{- g(x)}} \,
 \frac{\delta S_{\rm IF}[g^\pm]}
{\delta g^+_{ab}(x)} \right|_{g^\pm=g},
\label{s-t expect value}
\end{equation}
where the expectation value is taken in the $n$-dimensional
spacetime generalization of the state described by
$\hat{\rho}^{\rm \scriptscriptstyle S}(t_i)$. Therefore,
differentiating $S_{\rm eff}[g^\pm]$ in Eq. (\ref{ctpif}) with
respect to $g^+_{ab}$, and then setting
$g^+_{ab}=g^-_{ab}=g_{ab}$, we get the semiclassical Einstein
equation in $n$ dimensions. This equation is then renormalized by
absorbing the divergences in the regularized $\langle\hat
T^{ab}[g]\rangle$ into the bare parameters. Taking the limit
$n\to 4$ we obtain the physical semiclassical Einstein equation
(\ref{2.5}).


In the spirit of the previous derivation of the Einstein-Langevin
equation, we now seek a dynamical equation for a linear
perturbation $h_{ab}$ to the semiclassical metric $g_{ab}$,
solution of Eq. (\ref{2.5}). Strictly speaking if we use
dimensional regularization we must consider the $n$-dimensional
version of that equation; see Ref. \cite{MV1} for details. {}From
the results just described, if such an equation were simply a
linearized semiclassical Einstein equation, it could be obtained
from an expansion of the effective action $S_{\rm eff}[g+h^\pm]$.
In particular, since, from Eq. (\ref{s-t expect value}), we have
that
\begin{equation}
\langle \hat{T}^{ab}[g+h;x) \rangle =
\left. {2\over\sqrt{-\det (g\!+\!h)(x)}} \,
 \frac{\delta S_{\rm IF}
   [g\!+\!h^\pm]}{\delta h^+_{ab}(x)}
 \right|_{h^\pm=h},
\label{perturb s-t expect value}
\end{equation}
the expansion of $\langle \hat{T}^{ab}[g\!+\!h]\rangle $
to linear order in $h_{ab}$ can be obtained from an expansion of the
influence action $S_{\rm IF}[g+h^\pm]$ up to second order
in $h^{\pm}_{ab}$.

To perform the expansion of the influence action,
we have to compute the first and second order
functional derivatives of $S_{\rm IF}[g+h^\pm]$
and then set $h^+_{ab}\!=\!h^-_{ab}\!=\!h_{ab}$.
If we do so using the path integral representation
(\ref{path integral}), we can interpret these derivatives as
expectation values of operators.
The relevant second order derivatives are
\begin{eqnarray}
\left. {4\over\sqrt{- g(x)}\sqrt{- g(y)} } \,
 \frac{\delta^2 S_{\rm IF}[g+h^\pm]}
{\delta h^+_{ab}(x)\delta h^+_{cd}(y)}
 \right|_{h^\pm=h} \!\!
&=& -H_{\scriptscriptstyle \!
{\rm S}}^{abcd}[g;x,y)
-K^{abcd}[g;x,y)+
i N^{abcd}[g;x,y),      \nonumber \\
\left. {4\over\sqrt{- g(x)}\sqrt{- g(y)} } \,
 \frac{\delta^2 S_{\rm IF}[g^\pm]}
{\delta h^+_{ab}(x)\delta h^-_{cd}(y)}
 \right|_{h^\pm=h} \!\!
&=& -H_{\scriptscriptstyle \!
{\rm A}}^{abcd}
[g;x,y)
-i N^{abcd}[g;x,y),
\label{derivatives}
\end{eqnarray}
where
$$
N^{abcd}[g;x,y) \equiv
{1\over 2} \left\langle  \bigl\{
 \hat{t}^{ab}[g;x) , \,
 \hat{t}^{cd}[g;y)
 \bigr\} \right\rangle ,
\ \ \ \
H_{\scriptscriptstyle \!
{\rm S}}^{abcd}
[g;x,y) \equiv
{\rm Im} \left\langle {\rm T}^*\!\!
\left( \hat{T}^{ab}[g;x) \hat{T}^{cd}[g;y)
\right) \right\rangle \!,
$$
$$
H_{\scriptscriptstyle \!
{\rm A}}^{abcd}
[g;x,y) \equiv
-{i\over 2} \left\langle
\bigl[ \hat{T}^{ab}[g;x), \, \hat{T}^{cd}[g;y)
\bigr] \right\rangle \!,
\ \ \
K^{abcd}[g;x,y) \equiv
\left. {-4\over\sqrt{- g(x)}\sqrt{- g(y)} } \, \left\langle
\frac{\delta^2 S_m[g+h,\phi]}
{\delta h_{ab}(x)\delta h_{cd}(y)}
\right|_{\phi=\hat{\phi}}\right\rangle \!,
$$
with $\hat{t}^{ab}$ defined in Eq. (\ref{2.9}), $[ \; , \: ]$
denotes the commutator and $\{ \; , \: \}$ the anti-commutator.
Here we use a Weyl ordering prescription for the operators in the
last of these expressions the symbol ${\rm T}^*$ to denote the
following ordered operations: First, time order the field
operators $\hat{\phi}$ and then apply the derivative operators
which appear in each term of the product $T^{ab}(x) T^{cd}(y)$,
where $T^{ab}$ is the functional (\ref{2.3}). This ${\rm T}^{*}$
``time ordering'' arises because we have path integrals
containing products of derivatives of the field, which can be
expressed as derivatives of the path integrals which do not
contain such derivatives. Notice, from their definitions, that
all the kernels which appear in expressions (\ref{derivatives})
are real and also $H_{\scriptscriptstyle \!{\rm A}}^{abcd}$ is
free of ultraviolet divergences in the limit $n \to 4$.

{}From (\ref{s-t expect value}) and
(\ref{derivatives}), taking into account that
$S_{\rm IF}[g,g]=0$ and that
$S_{\rm IF}[g^-,g^+]=
-S^{ {\displaystyle \ast}}_{\rm IF}[g^+,g^-]$, we can write the
expansion for the influence action
$S_{\rm IF}[g\!+\!h^\pm]$ around a background
metric $g_{ab}$ in terms of the previous kernels.
Taking into account that
these kernels satisfy the symmetry relations
\begin{equation}
H_{\scriptscriptstyle \!{\rm S}}^{abcd}(x,y)=
H_{\scriptscriptstyle \!{\rm S}}^{cdab}(y,x),
\ \
H_{\scriptscriptstyle \!{\rm A}}^{abcd}(x,y)=
-H_{\scriptscriptstyle \!{\rm A}}^{cdab}(y,x),
\ \
K^{abcd}(x,y) = K^{cdab}(y,x),
\label{symmetries}
\end{equation}
and introducing the new kernel
\begin{equation}
H^{abcd}(x,y)\equiv
H_{\scriptscriptstyle \!{\rm S}}^{abcd}(x,y)
+H_{\scriptscriptstyle \!{\rm A}}^{abcd}(x,y),
\label{H}
\end{equation}
the expansion of $S_{\rm IF}$ can be finally written as
\begin{eqnarray}
S_{\rm IF}[g\!+\!h^\pm]
&=& {1\over 2} \int\! d^4x\, \sqrt{- g(x)}\:
\langle \hat{T}^{ab}[g;x) \rangle  \,
\left[h_{ab}(x) \right] \nonumber\\
&&-{1\over 8} \int\! d^4x\, d^4y\, \sqrt{- g(x)}\sqrt{- g(y)}\,
\left[h_{ab}(x)\right]
\left(H^{abcd}[g;x,y)\!
+\!K^{abcd}[g;x,y) \right)
\left\{ h_{cd}(y) \right\}  \nonumber  \\
&&
+{i\over 8} \int\! d^4x\, d^4y\, \sqrt{- g(x)}\sqrt{- g(y)}\,
\left[h_{ab}(x) \right]
N^{abcd}[g;x,y)
\left[h_{cd}(y) \right]+0(h^3),
\label{expansion 2}
\end{eqnarray}
where we have used the notation
\begin{equation}
\left[h_{ab}\right] \equiv h^+_{ab}\!-\!h^-_{ab},
\hspace{5 ex}
\left\{ h_{ab}\right\} \equiv h^+_{ab}\!+\!h^-_{ab}.
\label{notation}
\end{equation}
{}From Eqs.~(\ref{expansion 2}) and
(\ref{perturb s-t expect value})
it is clear that the imaginary part of the
influence action does not contribute to the perturbed
semiclassical Einstein equation (the expectation value of the
stress-energy tensor is real), however, as it depends on the noise kernel,
it contains information on the fluctuations of the operator
$\hat{T}^{ab}[g]$.

We are now in a position to carry out the derivation of the
semiclassical Einstein-Langevin equation. The procedure is well
known \cite{CH94,HuSin,CamVer96,thermal}: it consists of deriving
a new ``stochastic'' effective action from the observation that
the effect of the imaginary part of the influence action
(\ref{expansion 2}) on the corresponding influence functional is
equivalent to the averaged effect of the stochastic source
$\xi^{ab}$ coupled linearly to the perturbations $h_{ab}^{\pm}$.
This observation follows from the identity first invoked by
Feynman and Vernon for such purpose:
\begin{equation}
e^{-{1\over 8} \!\int\! d^4x\, d^4y \, \sqrt{- g(x)}\sqrt{- g(y)}\,
\left[h_{ab}(x) \right]\,
N^{abcd}(x,y)\, \left[h_{cd}(y)\right] }=
\int\! {\cal D}\xi \: {\cal P}[\xi]\, e^{{i\over 2} \!\int\! d^4x \,
\sqrt{- g(x)}\,\xi^{ab}(x)\,\left[h_{ab}(x) \right] },
\label{Gaussian path integral}
\end{equation}
where ${\cal P}[\xi]$ is the probability distribution
functional of a Gaussian stochastic tensor $\xi^{ab}$
characterized by the correlators (\ref{2.10})
with $N^{abcd}$ given by Eq.~(\ref{2.8}),
and where
the path integration measure is assumed to be a scalar under
diffeomorphisms of $({\cal M},g_{ab})$. The above identity follows
from the identification of the right-hand side of
(\ref{Gaussian path integral}) with the characteristic functional for
the stochastic field $\xi^{ab}$. The
probability distribution functional for $\xi^{ab}$ is explicitly
given by
\begin{equation}
{\cal P}[\xi]= {\rm det}(2\pi N)^{-1/2}
 \exp\left[-{1\over2}\!\int\! d^4x\, d^4y \, \sqrt{-g(x)}\sqrt{-g(y)}\,
 \xi^{ab}(x) \, N^{-1}_{abcd}(x,y)\, \xi^{cd}(y)\right].
\end{equation}

We may now introduce the {\it stochastic effective action} as
\begin{equation}
S^s_{\rm eff}[g+h^\pm,\xi] \equiv S_{g}[g+h^+]-S_{g}[g+h^-]+
S^s_{\rm IF}[g+h^\pm,\xi],
\label{stochastic eff action}
\end{equation}
where the ``stochastic'' influence action is defined as
\begin{equation}
S^s_{\rm IF}[g+h^\pm,\xi] \equiv {\rm Re}\, S_{\rm
IF}[g\!+\!h^\pm]+\! {1\over 2} \int\! d^4x \, \sqrt{-
g(x)}\,\xi^{ab}(x)\left[h_{ab}(x) \right]+ O(h^3). \label{eff
influence action}
\end{equation}
Note that, in fact, the influence functional can now be written as a
statistical average over $\xi^{ab}$:
$
{\cal F}_{\rm IF}[g+h^\pm]= \left\langle
\exp\left(i S^s_{\rm IF}[g+h^\pm,\xi]\right)
\right\rangle_{\! s}.
$
The stochastic equation of motion for $h_{ab}$ reads
\begin{equation}
\left.
\frac{\delta S^s_{\rm eff}[g+h^\pm,\xi]}{\delta h^+_{ab}(x)}
\right|_{h^\pm=h}=0,
\label{eq of motion}
\end{equation}
which is the Einstein-Langevin equation (\ref{2.11}); notice that only the
real part of $S_{IF}$ contributes to the expectation value
(\ref{perturb s-t expect value}). To be precise, we get
the regularized $n$-dimensional equations with the bare parameters,
and after renormalization we take the limit $n\to 4$ to obtain the
Einstein-Langevin equation in physical spacetime.

Before ending this section let us consider the causality property
of equation (\ref{eq of motion}). On general grounds causality is
guaranteed from the properties of the expectation values of
renormalized stress-energy tensor operators \cite{Wald94}. It is
illustrative, however, to check it explicitly in this case.
First, we note that in the expression (\ref{expansion 2}) of the
$S_{\rm IF}$ only  terms which involve the kernels $H^{abcd}$ and
$K^{abcd}$ may contain problems concerning causality. The kernel
$K^{abcd}$ is local and need not concern us. We want to show that
the nonlocal kernel $H^{abcd}$ leads to causal equations, as can
be seen from its structure. To simplify the proof let us assume a
quantum mechanical operator $\hat q(t)$, instead of the
stress-energy field operator \cite{RV99}, and suppress tensorial
indices, so that in this simplified problem the corresponding
kernel will be written as $H(t,t^\prime)$. Its general structure,
as follows from the definitions in (\ref{derivatives}), is
$H(t,t^\prime)= H_{\scriptscriptstyle \!{\rm S}}(t,t^\prime)+
H_{\scriptscriptstyle \!{\rm A}}(t,t^\prime)$, where $
H_{\scriptscriptstyle \!{\rm S}}(t,t^\prime)= {\rm Im}\langle
T(\hat q(t)\hat q(t^\prime))\rangle$, with $T$ denoting time
ordering, and $H_{\scriptscriptstyle \!{\rm A}}(t,t^\prime)=
-(i/2)\langle [\hat q(t),\hat q(t^\prime)]\rangle$. From this we
have that $H(t,t^\prime)= -i\langle [\hat q(t),\hat
q(t^\prime)]\rangle \theta(t-t^\prime)$. Since $H(t,t^\prime)$
appears in the equation of motion at time $t$ in a term such as
$\int dt^\prime H(t,t^\prime)h(t^\prime)$, where $h(t^\prime)$
plays the role of the metric perturbation, it is clear that the
nonlocal term depends on $h(t^\prime)$ for times $t^\prime<t$
only so that causality is guaranteed. When we go to the
stress-energy tensor field operator the proof is essentially the
same \cite{RV02a}: since $\hat T^{ab}(x)$ involves spacetime
derivatives acting on $\hat\phi(x)$ the corresponding $\theta$
functions in some of the terms will also carry derivatives but
these lead to delta functions which do not destroy the causality
property. Furthermore, time ordering is well defined in our
spacetime manifold, which is assumed to be globally hyperbolic
and thus time orientable.

\vskip 1cm
\centerline{\bf APPLICATIONS: Backreaction of Particle Creation }
\vskip .5cm

Particle creation from a strong or time-dependent gravitational
field as in cosmological \cite{cpc} and black hole
\cite{bhpc,bhpc1} spacetimes was studied in the late 60's to the
70's. The effects of particle creation, vacuum polarization and
other quantum processes back acting on the background spacetime
constitutes what is known as the backreaction problem.
Backreaction become increasingly important when the energy
reaches the Planck scale, as in the early universe \cite{cpcbkr}
and at the final stages of black hole evolution
\cite{HI,Bardeen,York}. Investigation of backreaction problems
started with the study of regularization of the stress energy
tensor in curved spacetimes in the mid-70's, followed by in-depth
calculations of backreaction of particle creation in cosmological
spacetimes, such as possible removal of the cosmological
singularity by  the trace anomaly, and the damping of anisotropy
in Bianchi universes by particle creation.

The most significant developments in the implementation and
physical aspects of the backreaction problems exemplified in
cosmological spacetimes in the 80's are perhaps the introduction
of an effective action which yields real and causal equations of
motion. (For the axiomatic theoretical aspects, see, e.g., Kay
and Wald, Flanagan and Wald \cite{KayWal,FlaWal}.) This is known
as the closed-time-path (CTP) or the Schwinger-Keldysh method
\cite{ctp}. From this one can identify dissipative effects in an
unambiguous manner and in the true statistical mechanical sense.
In the 90's, the most significant development was perhaps the
introduction of quantum open systems concepts \cite{qos} and the
influence functional method \cite{FeyVer,ifqbm}, which enables
one to identify the origin of quantum noise and recognize the
importance of fluctuations. We give an example of the calculation
of the backreaction of cosmological particle creation to
illustrate the use of these methods which rest at the base of
stochastic gravity theory.

\section{Cosmological Backreaction Problems}
\label{sec:cosbkrn}

As a canonical example of cosmological backreaction we consider
in this section a massless conformally coupled quantum scalar
field on a weakly perturbed spatially flat
Friedmann-Robertson-Walker (FRW) spacetime and derive the
semiclassical Einstein-Langevin equation for the metric
perturbations off this spacetime. These equations were obtained
in Ref.~\cite{CamVer96} following the derivation of the CTP
effective action for this problem in Ref.~\cite{CamVer94}.
Einstein-Langevin equations had been previously derived in
Ref.~\cite{HuSin} for small anisotropies conformally coupled to
massless fields in a spatially homogeneous background working in
the framework of quantum cosmology, and in Ref.~\cite{HM3} for
the scale factor in a spatially flat universe due to the coupling
of different quantum scalar fields. The connection between the
CTP effective action and the influence functional had been
noticed in Ref.~\cite{CH94} in the semiclassical gravity context.

To derive the Einstein-Langevin equation we can compute the
influence action $S_{\rm IF}$, defined in Eq. (\ref{expansion
2}), which is equivalent to evaluating the different kernels
introduced after Eq. (\ref{derivatives}). This can be done
directly from the expressions for the kernels in terms of
products of the Feynman and the Wightman propagators for the
scalar field in the background metric which were derived in
Refs.~\cite{MV1} and \cite{MV2}, or it can be done by an explicit
evaluation of the path integrals which define the influence action
in Eqs.~(\ref{path integral}) and (\ref{influence functional}).
We follow this second, more direct, route in this section.

The metric of our spacetime is given by
\begin{equation}
   \tilde g_{\mu\nu}(x)=a^2(\eta)
                                  \left(\eta_{\mu\nu}+h_{\mu\nu}(x)
                                  \right),
\label{eq:FRWL}
\end{equation}
where $\eta_{\mu\nu}=\mbox{\rm diag}(-1,+1,\cdots,+1)$, $a(\eta)
\equiv e^{\omega(\eta)}$ is the scale factor, $\eta$ is the
conformal time which is related to the cosmological time $t$ by
$dt=a d\eta$, $h_{\mu\nu}(x)$ is a symmetric tensor which
represents arbitrary small metric perturbations,  and we work in
an $n$ dimensional spacetime in preparation for dimensional
regularization.

The classical action for a free massless conformally coupled real
scalar field $\Phi(x)$ is given in $n$-dimensions by
\begin{equation}
S_m[\tilde g_{\mu\nu},\Phi]=
        -{1\over2}\int d^nx\sqrt{-\tilde g}
            \left[\tilde g^{\mu\nu}\partial_\mu\Phi\partial_\nu\Phi
                 +\xi(n)\tilde R\Phi^2
            \right],
\label{4.0}
\end{equation}
where $\xi(n)=(n-2)/[4(n-1)]$, and $\tilde R$ is the Ricci scalar
for the metric $\tilde g_{\mu\nu}$. Because of the conformal
coupling $\xi(n)$ we can define a new field $\phi(x)$ and a new
metric $g_{\mu\nu}$
as
\begin{equation}
\phi(x)\equiv e^{(n/2-1)\omega(\eta)}\Phi(x),\ \
g_{\mu\nu}(x)\equiv \eta_{\mu\nu}+h_{\mu\nu}(x),
\label{5.1}
\end{equation}
so that $g_{\mu\nu}$ is conformally related
to $\tilde g_{\mu\nu}$.
After one integration by parts
the classical action (\ref{4.0}) takes the form
\begin{equation}
   S_m[g_{\mu\nu},\phi] =
        -{1\over2}\int d^nx\sqrt{- g}
            \left[g^{\mu\nu}\partial_\mu\phi\partial_\nu\phi
                 +\xi(n) R\phi^2
            \right]
\end{equation}
which is the action for a free massless conformally coupled real
scalar field $\phi(x)$ in a spacetime with metric $g_{\mu\nu}$,
{\it i.e.} a nearly flat spacetime. Although the physical field
is $\Phi(x)$ the fact that it is related to the field $\phi(x)$
by a power of the conformal factor implies that a positive
frequency mode of the field $\phi(x)$ in flat spacetime will
correspond to a positive frequency mode in the conformally related
space; these modes define the conformal vacuum. Thus quantum
effects such as particle creation will be due to the breaking of
conformal flatness which, in this case, is produced by the
perturbations $h_{\mu\nu}(x)$. Expanding the above action in
terms of these perturbations, after integrations by parts we have
\begin{equation}
   S_m[g_{\mu\nu},\phi]={1\over2}\int d^nx\,
                   \phi[\Box +V^{(1)}+V^{(2)}+\dots
                   ]\phi,
   \label{eq:matter action}
\end{equation}
where, the operators $V^{(1)}$ and $V^{(2)}$ are defined as
\begin{eqnarray}
      V^{(1)}(x)&=&-\bar h^{\mu\nu}\partial_\mu\partial_\nu
                -\left(\partial_\mu \bar h^{\mu\nu}\right)\partial_\nu
                -\xi(n)R^{(1)},
\nonumber\\
       V^{(2)}(x)&=& \hat h^{\mu\nu}\partial_\mu\partial_\nu
                +\left(\partial_\mu \hat h^{\mu\nu}\right)\partial_\nu
                -\xi(n)\left(R^{(2)}+{1\over2}hR^{(1)}\right),
\label{operatorsV}
\end{eqnarray}
where
\begin{eqnarray}
\bar h_{\mu\nu}&\equiv&h_{\mu\nu}-{1\over2}h\eta_{\mu\nu},
\nonumber\\
\hat h_{\mu\nu}&\equiv& {h_\mu}^\alpha h_{\alpha\nu}-{1\over2}
hh_{\mu\nu}+{1\over8}h^2\eta_{\mu\nu}-{1\over4}h_{\alpha\beta}
h^{\alpha\beta}\eta_{\mu\nu},
\label{metric perturbations}
\end{eqnarray}
and $R^{(1)}$ and $R^{(2)}$
are the first and second order terms, respectively, in the metric
perturbations of the scalar curvature.

To the classical action for the matter fields $S_m$ we have to add
the action of the physical metric $\tilde g_{\mu\nu}$:
$S_g[\tilde g_{\mu\nu}]$. As it was emphasized in Sec.~\ref{sec2}
in order to renormalize the effective action we need to add
appropriate terms quadratic in the Riemann tensor. In this case,
using dimensional regularization the only terms needed for
renormalization are:
\begin{eqnarray}
     S_g[\tilde g_{\mu\nu}]&\equiv & \int d^nx
        \sqrt{-\tilde g(x)} \left\{ \frac{1}{16\pi G_N}\tilde R(x)\right.
        \nonumber \\
     & &\hskip1cm
        +\frac{\mu^{n-4}}{2880\pi^2(n-4)}\left[ \tilde R_{\mu\nu\alpha\beta}
        (x)\tilde R^{\mu\nu\alpha\beta}(x)-\tilde R_{\mu\nu}(x)
        \tilde R^{\mu\nu}(x)\right]\left. {\atop }\right\}.
\label{4.1}
\end{eqnarray}
where $G_N$ denotes Newton's gravitational constant and $\mu$ is
a mass renormalization parameter. Note that in $4$-dimensions
$\tilde R_{\mu\nu\alpha\beta}
        (x)\tilde R^{\mu\nu\alpha\beta}(x)-\tilde R_{\mu\nu}(x)
        \tilde R^{\mu\nu}(x)=(3/2)\tilde C_{\mu\nu\alpha\beta}
        (x)\tilde C^{\mu\nu\alpha\beta}(x)$ but this is not true when
$n\not= 4$, for that reason we have to use the previous combination
of Riemann and Ricci tensors instead of the Weyl tensor; furthermore we can add
a term proportional to $\tilde R^2(x)$ in the gravitational
action with an arbitrary coefficient,
but since this term is not needed for renormalization
we do not introduce it here.

To compute the influence action
$S_{\rm IF}$ from Eqs.~(\ref{path integral}) and (\ref{influence functional})
we have to introduce two scalar fields $\phi_+(x)$ and $\phi_-(x)$
which coincide at some future time $t_f$,
$\phi_+(t_f)=\phi_-(t_f)$, and which evolve in two
different geometries given by $h^+_{\mu\nu}$ and $h^-_{\mu\nu}$
such that $h^+_{\mu\nu}(t_f)=h^-_{\mu\nu}(t_f)$.
The standard Gaussian path integral computation leads to
\begin{equation}
S_{\rm IF}[h^\pm_{\mu\nu}]=-\frac{i}{2}{\rm Tr}(\ln G),
\end{equation}
where G is a $2\times 2$ matrix propagator for the fields $\phi_+(x)$
and $\phi_-(x)$ that may be obtained from the action
(\ref{eq:matter action}) for the (+) field minus the same action for
the (--) field. This propagator cannot be found exactly but it
can be obtained perturbatively in powers of the metric perturbations. The
unperturbed matrix propagator, $G^0$, is the inverse of the
kinetic operator ${\rm diag}(\Box-i\epsilon, -(\Box +i\epsilon))$,
where we introduced the usual prescriptions for the vacuum state. It has the
following components:
$G^0_{++}=\Delta_F$, $G^0_{--}=-\Delta_D$, $G^0_{+-}=-\Delta^+$ and
$G^0_{-+}=\Delta^-$, where $\Delta_F$ and $\Delta_D$ are the
Feynman and Dyson propagators, respectively, and $\Delta^\pm$ are the
Wightman functions.

Up to second order in the metric perturbations we get,
\begin{eqnarray}
   S_{\rm IF}[h^\pm_{\mu\nu}]
        &\simeq &
                 -{i\over2}{\rm Tr} (\ln G^0)
            \nonumber \\
         & &+{i\over2}{\rm Tr}
                    \left[\atop\right.
                         V^{(1)}_+G^0_{++}-V^{(1)}_-G^0_{--}
                        +V^{(2)}_+G^0_{++}-V^{(2)}_-G^0_{--}
            \nonumber \\
         & &\hskip1cm
                        -{1\over2}V^{(1)}_+G^0_{++}V^{(1)}_+G^0_{++}
                        -{1\over2}V^{(1)}_-G^0_{--}V^{(1)}_-G^0_{--}
                        +V^{(1)}_+G^0_{+-}V^{(1)}_-G^0_{-+}
                    \left.\atop\right],
         \label{eq:formal CTP effective action}
\end{eqnarray}
where we have defined $V^{(i)}_+\equiv V^{(i)}_{++}$ and
$V^{(i)}_-\equiv -V^{(i)}_{--}$ ($i=1,2$), since the operators
$V^{(i)}$ are obviously diagonal. The first trace term is
independent of the metric perturbations, its divergences are
cancelled by terms which lead to the conformal anomaly. The
tadpole terms of type ${\rm Tr}(VG)$ involve $n$-dimensional
integrals which are identically zero in dimensional
regularization. Thus, all calculation reduces to the computation
of the three terms of type ${\rm Tr}(VGVG)$. The detailed
evaluation of these terms is given in Ref.~\cite{CamVer94}.

After dimensional regularization  of the divergent terms and
renormalization with the action of the gravitational action
(\ref{4.1}), we obtain the renormalized effective action,
(\ref{ctpif}), for the gravitational field. Up to second order it
can be written as
\begin{equation}
S_{\rm eff}[{\tilde g}^\pm_{\mu\nu}]=S^{R}_{g}[{\tilde g}^+_{\mu\nu}]
           -S^{R}_{g}[{\tilde g}^-_{\mu\nu}]+S_{\rm IF}^R[h_{\mu\nu}^\pm],
\label{eq:renormalized CTP}
\end{equation}
where the renormalized influence functional is given by
\begin{eqnarray}
  S_{\rm IF}^R[h_{\mu\nu}^\pm] &=&
        -\frac{1}{8}
    \int d^4xd^4y[h_{\mu\nu}(x)]H^{\mu\nu\alpha\beta}(x,y;\mu)\{h_{\alpha\beta}(y)\}
           \nonumber \\
        & &+\frac{i}{8}
    \int d^4xd^4y [h_{\mu\nu}(x)]N^{\mu\nu\alpha\beta}(x,y)[h_{\alpha\beta}(y)],
   \label{4.2}
\end{eqnarray}
and $S^R_g[\tilde g^\pm_{\mu\nu}]$ are terms coming from the
gravitational action (\ref{4.1}) after renormalization:
\begin{eqnarray}
   S^{R}_g[{\tilde g}_{\mu\nu}]
        &=&\int d^4x\sqrt{-\tilde g(x)}
              \left[ \frac{\tilde R(x)}{16\pi G_N}
                    -\frac{1}{12} \frac{1}{2880\pi^2}\tilde R^2(x)
              \right]
           \nonumber \\
        & &+\frac{1}{1440\pi^2} \int d^4x\sqrt{-g(x)}
              \left[ G^{\mu\nu}(x)\omega_{;\mu}\omega_{;\nu}
                    +\Box_g\omega(\omega_{;\nu}\omega^{;\nu})
                    +{1\over 2}(\omega_{;\mu} \omega^{;\mu})^2
              \right]
           \nonumber \\
        & &+\frac{1}{2880\pi^2}\int d^4x\sqrt{-g(x)}
              \left[ R_{\mu\nu\alpha\beta}(x)
                     R^{\mu\nu\alpha\beta}(x)
                    -R_{\mu\nu}(x)R^{\mu\nu}(x)
              \right] \omega (x).
   \label{eq:renormalized gravitational action}
\end{eqnarray}
Here we have used the notation of Eq.~(\ref{notation}) and the
kernels $H^{\mu\nu\alpha\beta}$ and $N^{\mu\nu\alpha\beta}$ are
given by
\begin{equation}
H^{\mu\nu\alpha\beta}(x,y;\mu)=\frac{2}{3}F^{\mu\nu\alpha\beta}_xH(x-y;\mu),
\label{kernelH1}
\end{equation}
where $F^{\mu\nu\alpha\beta}_x$ is the differential operator
\begin{equation}
F^{\mu\nu\alpha\beta}_x\equiv 3F^{\mu(\alpha}_x F^{\beta)\nu}_x-
F^{\mu\nu}_x F^{\alpha\beta}_x,\ \ \
F^{\mu\nu}_x\equiv \eta^{\mu\nu}\Box_x-\partial^\mu_x \partial^\nu_x,
\label{operatorF}
\end{equation}
and
\begin{equation}
H(x;\mu)=\frac{1}{1920\pi^2}\int \frac{d^4 p}{(2\pi)^4}e^{ip\cdot
x} \left[ \ln\frac{|p^2|}{\mu^2}-i\pi \,{\rm
sgn}(p^0)\theta(-p^2)\right]. \label{kernelH2}
\end{equation}
The noise kernel is given by
\begin{equation}
N^{\mu\nu\alpha\beta}(x,y)=\frac{2}{3}F^{\mu\nu\alpha\beta}_xN(x-y),
\label{kernelN1}
\end{equation}
where
\begin{equation}
N(x)=\frac{1}{1920\pi}\int \frac{d^4 p}{(2\pi)^4}e^{ip\cdot x}\theta(-p^2).
\label{kernelN2}
\end{equation}
Note that terms with and without tilde refer to tensors obtained with metrics
$\tilde g_{\mu\nu}$ and $g_{\mu\nu}$, respectively.

We are now in a position to derive the Einstein-Langevin equation.
Following section \ref{sec3} we can introduce the stochastic
effective action
\begin{equation}
S_{\rm eff}^s[{\tilde g}^\pm_{\mu\nu},\xi]=S^{R}_{g}[{\tilde g}^+_{\mu\nu}]
           -S^{R}_{g}[{\tilde g}^-_{\mu\nu}]+S_{\rm IF}^{R,s}[h_{\mu\nu}^\pm,\xi],
\label{stochastic-action}
\end{equation}
where the renormalized stochastic influence action is given by
\begin{equation}
S_{\rm IF}^{R,s}[h_{\mu\nu}^\pm,\xi]={\rm Re}\,S_{\rm
IF}^{R}[h_{\mu\nu}^\pm] + \frac{1}{2}\int d^4 x
\,\xi^{\mu\nu}(x)[h_{\mu\nu}(x)], \label{stochastic-influence}
\end{equation}
with the Gaussian stochastic field $\xi^{\mu\nu}$ defined by
$\langle\xi^{\mu\nu}(x)\rangle_s =0$ and
\begin{equation}
\langle \xi^{\mu\nu}(x)\xi^{\alpha\beta}(y)\rangle_s=N^{\mu\nu\alpha\beta}(x,y).
\label{correlations}
\end{equation}
The Einstein-Langevin equation can be obtained by functional
derivation according to Eq.~(\ref{eq of motion}). Note that we
have to take the derivative with respect to the physical metric
$\tilde g_{\mu\nu}$, and we use that for an arbitrary functional
$A[\tilde g_{\mu\nu}]$,
\begin{equation}
     \frac{\delta A [\omega , g_{\mu\nu}]} {\sqrt{-g}\delta g_{\mu\nu}}=
     e^{6\omega}
     \frac{\delta A [\tilde g_{\mu\nu}]}
     {\sqrt{-\tilde g}\delta \tilde g_{\mu\nu}}.
\end{equation}
The final result is:
\begin{eqnarray}
   & &
      e^{6\omega}\left[-{1\over 8\pi G_N}
                          \left( \tilde G^{\mu\nu}_{(0)}
                                +\tilde G^{\mu\nu}_{(1)}
                          \right)
                       -\frac{1}{6}\frac{1}{2880\pi^2}
                          \left( \tilde B^{\mu\nu}_{(0)}
                                +\tilde B^{\mu\nu}_{(1)}
                          \right)\right.
         \nonumber\\
                       & &\hskip0.8cm\left.+\frac{1}{2880\pi^2}
                          \left( \tilde H^{\mu\nu}_{(0)}
                                +\tilde H^{\mu\nu}_{(1)}
                          \right)
                       -\frac{1}{1440\pi^2}\tilde R^{(0)}_{\alpha\beta}
                              \tilde C^{\mu\alpha\nu\beta}_{(1)}
                 \right]
      \nonumber \\
   & &\hskip1.2cm
      -\frac{3}{720\pi^2}
        (C^{\mu\alpha\nu\beta}_{(1)}\omega )_{,\alpha\beta}
              -\int d^4y A^{\mu\nu}_{(1)}(y)\mbox{\rm H}(x-y;\mu)
        +\xi^{\mu\nu}
      =O(h^2_{\mu\nu}),
   \label{eq:semiclassical equations bis}
\end{eqnarray}
where the $(0)$ and $(1)$ sub-indices refer to the zero and first
orders terms, respectively, in the metric perturbation
$h_{\mu\nu}$. The tensors $A^{\mu\nu}(x)$ and $B^{\mu\nu}(x)$ are
given by Eqs.~(\ref{2.7a}) and (\ref{2.7b}), respectively, and
$H^{\mu\nu}(x)$ is given by
\begin{equation}
   H^{\mu\nu}=
               -R^{\mu\alpha}{R_\alpha}^\nu
               +{2\over3}RR^{\mu\nu}
               +{1\over2}g^{\mu\nu}R_{\alpha\beta}R^{\alpha\beta}
               -{1\over4}g^{\mu\nu}R^2.
\end{equation}

When comparing the Einstein-Langevin Eq.~(\ref{eq:semiclassical
equations bis}) with Eq.~(\ref{2.11}) we recall that in our
renormalization scheme [see Eq.~(\ref{4.1})], we have implicitly
fixed the arbitrary parameter $\beta$,  whereas the parameter
$\alpha$ appears related to the parameter $\mu$. In fact, if we
change $\mu$ by $\mu^\prime$ in the kernel (\ref{kernelH2}) we
have
\begin{equation}
H(x-y;\mu)=H(x-y;\mu^\prime)
+\frac{1}{1920\pi^2}\delta^{(4)}(x-y)\ln \frac{\mu^2}{\mu^{\prime\, 2}},
\label{kernel relations}
\end{equation}
therefore the arbitrariness of $\alpha$ corresponds to that of
the renormalization parameter $\mu$. To be specific, one must
assume that $\alpha(\mu)$ and that if we change $\mu$ by
$\mu^\prime$ also $\alpha$ changes appropriately so that the
physical parameters in the Einstein-Langevin equation do not
change. Note that $\tilde A^{\mu\nu}_{(0)}=0$ because $\tilde
A^{\mu\nu}$ is obtained by taking the functional derivative of an
action density proportional to the square of the Weyl tensor, see
(\ref{2.7a}), but the conformal symmetry is broken only by the
metric perturbations $h_{\mu\nu}$, at zero order in the
perturbation the physical metric is conformally flat and the Weyl
tensor is zero. Since the parameters $\alpha$ and $\beta$ as well
as the gravitational constant can only be fixed by experiment we
will introduce a (rescaled) parameter $\beta$ in our final
equations.

Let us write the Einstein-Langevin
Eq.~(\ref{eq:semiclassical equations bis}) with the effective source term
on the right-hand side as
\begin{equation}
   \tilde G^{\mu\nu}=
         8\pi G_N\left( \langle T^{\mu\nu} \rangle
                +e^{-6\omega}\xi^{\mu\nu}
                     \right),
   \label{eq:effective stress tensor}
\end{equation}
where $\langle T^{\mu\nu} \rangle$ is the vacuum
expectation value of the stress-energy tensor of the quantum field
up to first order in $h_{\mu\nu}$. It
is given by
\begin{eqnarray}
   \langle T^{\mu\nu}_{(0)} \rangle
        &=&
            \frac{1}{2880\pi^2} \left[\tilde H^{\mu\nu}_{(0)}
    -\frac{\beta}{6} \tilde B^{\mu\nu}_{(0)}\right]
           \nonumber \\
   \langle T^{\mu\nu}_{(1)} \rangle
        &=&
  \frac{1}{2880\pi^2} \left[ \atop \right.
                         \tilde H^{\mu\nu}_{(1)}
         -\frac{\beta}{6} \tilde B^{\mu\nu}_{(1)}
                          -2\tilde R^{(0)}_{\alpha\beta}
                          \tilde C^{\mu\alpha\nu\beta}_{(1)}
              \nonumber \\
          & &\hskip2cm
                        -12e^{-6\omega}
                             ( C^{\mu\alpha\nu\beta}_{(1)}
                                       \omega
                                     )_{,\alpha\beta}\left. \atop \right]
                        -e^{-6\omega}\int d^4y A^{\mu\nu}_{(1)}(y)
                                      \mbox{\rm H}(x-y;\mu),
   \label{eq:quantum stress tensor}
\end{eqnarray}
where the rescaled $\beta$ parameter has been introduced. This
tensor was first derived in Ref.~\cite{CamVer94} using the CTP
formalism, but it had been derived by other means in
Refs.~\cite{horowitz-wald}. To summarize,  the stochastic equation
(\ref{eq:effective stress tensor}) is the semiclassical
Einstein-Langevin equation for weakly inhomogeneous perturbations
on spatially-flat FRW spacetimes with a conformally coupled
massless scalar field. Here, following Ref.~\cite{CamVer96} we
have derived a stochastic correction to the vacuum expectation
value of the quantum field stress-energy tensor (\ref{eq:quantum
stress tensor}), which accounts for the noise associated with the
fluctuations of the stress-energy tensor on the homogeneous
background spacetime.

We can recover the trace anomaly result
\cite{BirDav}. In fact, to first order in $h_{\mu\nu}$ we have:
\begin{eqnarray}
    \langle {T^\mu}_{\mu} \rangle
                   &=& \langle T^{\ \mu}_{(0)\mu} \rangle
                       +\langle T^{\ \mu}_{(1)\mu} \rangle
                       +O(h^2_{\mu\nu}) \nonumber \\
                   &=& \frac{1}{2880\pi^2}\left[\beta\Box_{\tilde g}\tilde R
                       +\left( \tilde R^{\mu\nu}\tilde R_{\mu\nu}
                       -{1\over 3}\tilde R^2\right)\right]
                       +O(h^2_{\mu\nu}).
\end{eqnarray}
Note that due to the conformal invariance of $\int d^4x\sqrt{-g}\tilde
C_{\mu\nu\alpha\beta}
\tilde C^{\mu\nu\alpha\beta}$ the tensor $\tilde A^{\mu\nu}$ is traceless.

As we have seen in Sec.~\ref{sec2} the stochastic correction to
the stress-energy tensor is traceless for conformal fields and has
vanishing divergence to first order in the metric perturbations.
These properties can be easily checked in this case. That
$\langle \xi^\mu_\mu\rangle_s=0$ follows directly from
Eq.~(\ref{correlations}) and (\ref{kernelN1}) by noticing that
$\langle N^{\mu\ \alpha\beta}_{\ \mu}(x,y)\rangle=0$ as a
consequence of $F^{\mu (\alpha}F^{\beta )}_{\ \ \mu}=\Box
F^{\alpha\beta}$ and $F^{\mu}_{\ \mu}=3\Box$. Furthermore, using
${\tilde g}_{\mu\nu}=e^{2\omega}g_{\mu\nu}$ and that
$\xi^{\mu\nu}$ is symmetric and traceless, it is easy to see that
${\tilde\nabla}_\nu\left(e^{-6\omega}\xi^{\mu\nu}\right)=
e^{-6\omega}\nabla_\nu \xi^{\mu\nu}$. Then, from Eq.
(\ref{correlations}) and the symmetries of $\xi$, we obtain that
$\nabla_\nu \xi^{\mu\nu}=O(h_{\mu\nu})$. It is thus consistent to
write this term on the right-hand side of Einstein equations and
consider it as a correction of order higher than $\langle
T^{\mu\nu}_{(0)}\rangle$ (note that ${\tilde\nabla}_\nu \langle
T^{\mu\nu}_{(0)}\rangle=O(h^2_{\mu\nu})$).

Taking the mean value of the Einstein-Langevin equation with
respect to the tensor field source $\xi^{\mu\nu}$ with Gaussian
probability distributions (\ref{correlations}), we obtain the
semiclassical Einstein equation. This equation can be used to
study the stability of the zero order semiclassical
Eq.~(\ref{eq:effective stress tensor}):
\begin{equation}
\tilde G^{\mu\nu}_{(0)}=
         8\pi G_N \langle T^{\mu\nu}_{(0)}\rangle.
\label{zero order}
\end{equation}
If we take the value $\beta=1$, which is the value obtained in
our renormalization scheme, this stress-energy tensor $\langle
T^{\mu\nu}_{(0)}\rangle$ for a scalar field in a conformally flat
spacetime exactly agrees with that found by other techniques
\cite{BirDav}. But in general, as remarked earlier, the
coefficient $\beta$ is arbitrary and should be determined by
experiment. On the other hand since the tensor $\tilde
A^{\mu\nu}_{(0)}=0$ there is no $\alpha$ coefficient [see Eq.
(\ref{2.11})]. Moreover, the coefficient of the tensor $\tilde
H^{\mu\nu}_{(0)}$ is fixed, its value differs for different types
and number of fields, it has been computed here for a scalar
field. Note that the tensor $\tilde H^{\mu\nu}$ is conserved only
in conformally flat spacetimes and cannot be obtained by
functional derivation of a geometrical term in the action, so
that it plays a very different role than the tensors $\tilde
A^{\mu\nu}$ and $\tilde B^{\mu\nu}$. Equation (\ref{zero order})
can be solved to find the conformal factor $\omega(\eta)$, it was
shown by Starobinsky \cite{Starobinsky80,Vilenkin85} that this
equation describes the so-called trace anomaly driven inflation.
Note that we can use the Einstein-Langevin equation
(\ref{eq:effective stress tensor}) to compute the two-point
correlations for the metric perturbations induced by the
stress-energy tensor fluctuations. These correlations are
relevant for the generation of primordial inhomogeneities in this
inflationary scenario \cite{Hawking01}. This problem has been
treated recently from the stochastic gravity perspective in
Ref.~\cite{RV02a} for inflationary models driven by scalar fields.

It is interesting to notice also that the imaginary term in the
regularized influence action (\ref{4.2}) can be written after an
integration by parts in the following alternative form
~\cite{CamVer96}
\begin{equation}
{\rm Im} \,S_{IF}^R[h^\pm_{\mu\nu}]=
 \frac{1}{2}
    \int d^4xd^4y [C_{\mu\nu\alpha\beta}(x)]N(x-y)[C^{\mu\nu\alpha\beta}(y)],
\end{equation}
which shows that the noise couples to the conformal tensor. Here
we have used the notation (\ref{notation}) for the square
brackets. The fact that the stochastic source couples to the
conformal tensor is not a surprise. For a conformal quantum field
non trivial quantum effects are a consequence of breaking the
conformal symmetry of the spacetime, which is characterized by
the conformal tensor. For instance, it is known that the
probability density of pair creation in this case \cite{CesVer90},
or in the presence of small anisotropy \cite{HuSin}, is
determined by the square of the Weyl tensor. Thus, as it has been
shown in Refs.~\cite{CH94,MV1} there is a direct relation between
particle creation and noise.

\section{Black Hole Backreaction Problem}
\label{sec:bhbkrn}

The celebrated Hawking effect of particle creation from black
holes is constructed from a quantum field theory in curved
spacetime (QFTCST) framework. The oft-mentioned `black hole
evaporation' referring to the reduction of the mass of a black
hole due to particle creation must entail backreaction
considerations. Backreaction of Hawking radiation
\cite{HI,Bardeen,York,HK,HKY,AHWY} could alter the evolution of
the background spacetime and change the nature of its end state,
more drastically so for Planck size black holes. Because of the
higher symmetry in cosmological spacetimes, backreaction studies
of processes therein have progressed further than the
corresponding black hole problems, which to a large degree is
still concerned with finding the right approximations for the
regularized energy momentum tensor \cite{JMO,MPP,AHS,AHL,HLA}
\footnote{The latest important work is that of Hiscock, Larson
and Anderson \cite{HLA} on backreaction in the interior of a
black hole, where one can find a concise summary of earlier work.
To name a few of the important landmarks in this endeavor (this
is adopted from \cite{HLA}), Howard and Candelas
\cite{HowCan,How} have computed the stress-energy of a
conformally invariant scalar field in the Schwarzschild geometry.
Jensen and Ottewill \cite{JenOtt} have computed the vacuum
stress-energy of a massless vector field in Schwarzschild.
Approximation methods have been developed by Page, Brown, and
Ottewill \cite{Page82,BroOtt,PBO} for conformally invariant
fields in Schwarzschild spacetime, Frolov and Zel'nikov
\cite{FZ1} for conformally invariant fields in a general static
spacetime, Anderson, Hiscock and Samuel \cite{AHS} for massless
arbitrarily coupled scalar fields in a general static spherically
symmetric spacetime. Furthermore the DeWitt-Schwinger
approximation has been derived by Frolov and
Zel'nikov\cite{FZ82,FZ84} for massive fields in Kerr spacetime,
Anderson Hiscock and Samuel \cite{AHS} for a general (arbitrary
curvature coupling and mass) scalar field in a general static
spherically symmetric spacetime and have applied their method to
the Reissner-Nordstr\"{o}m geometry \cite{AHL}.} for even the
simplest spacetimes such as the spherically symmetric family
including the important Schwarzschild metric. Though arduous and
demanding, the effort continues on because of the importance of
backreaction effects of Hawking radiation in black holes. They
are expected to address some of the most basic issues such as
black hole thermodynamics \cite{bhpc1,BekHaw,smbhent,strbhent} and
the black hole end-state and information loss puzzles
\cite{Pagereview}.

Here we wish to address the black hole backreaction problem with
new insights provided by stochastic semiclassical gravity (SSG).
(For the latest developments see reviews, e.g.,
\cite{Banff,stogra,HVErice,HVLivRev}). It is not our intention to
seek better approximations for the regularized energy momentum
tensor, but to point out new ingredients lacking in the existing
framework based on semiclassical gravity (SCG). In particular one
needs to consider both the dissipation and the fluctuations
aspects in the back reaction of particle creation or vacuum
polarization.

In a short note \cite{Vishu} Raval, Sinha and one of us (HRS)
discussed the formulation of the problem in this new light,
commented on some shortcomings of existing works, and sketched the
strategy \cite{SRH} behind our own approach to the black hole
fluctuations and backreaction problem. Here we focus only on the
class of quasi-static black holes, leaving the more demanding
dynamical collapse problem to a later exposition. Thus we only
address the first set of major issues mentioned above.

{}From the new perspective provided by statistical field theory
and stochastic gravity, it is not difficult to postulate that
backreaction effect is the manifestation of a fluctuation-
dissipation relation (FDR) \cite{FDR}. This was first conjectured
by Candelas and Sciama \cite{CanSci} for a dynamic Kerr black hole
emitting Hawking radiation, and Mottola \cite{Mottola} for a
static black hole (in a box) in quasi-equilibrium with its
radiation via linear response theory (LRT) \cite{LRT}. While the
FDR in a LRT captures the response of the system (e.g.,
dissipation of the black hole) to the environment (in these cases
the matter field) linear response theory (in the way it is
commonly presented in statistical thermodynamics)  cannot provide
a full description of self-consistent backreaction on at least two
counts: First, because it is usually based on the assumption of a
specified background spacetime (static in this case) and state
(thermal) of the matter field(s)(e.g., \cite{Mottola}). The
spacetime and the state of matter should be determined in a
self-consistent manner by their dynamics and mutual influence.
Second, the fluctuation part represented by the noise kernel is
amiss (e.g., \cite{AMM02}) This is also a problem in the FDR
proposed by Candelas and Sciama \cite{CanSci} (see below). As
will be shown in an explicit example later, the back reaction is
intrinsically a dynamic process. The Einstein-Langevin equation
in stochastic gravity overcomes both of these deficiencies.

For Candelas and Sciama \cite{CanSci}, the classical formula they
showed relating the dissipation in area linearly to the squared
absolute value of the shear amplitude is suggestive of  a
fluctuation-dissipation relation. When the gravitational
perturbations are quantized (they choose the quantum state to be
the Unruh vacuum) they argue that it approximates a flux of
radiation from the hole at large radii. Thus the dissipation in
area due to the Hawking flux of gravitational radiation is
allegedly related to the quantum fluctuations of gravitons.
HRS's criticism \cite{Vishu} is that their's is not an FDR in the
truly statistical mechanical sense because it does not relate
dissipation of a certain quantity (in this case, horizon area) to
the fluctuations of {\it the same quantity}. To do so would
require one to compute the two point function of the area, which,
being a four-point function of the graviton field, is related to
a two-point function of the stress tensor. The stress tensor is
the true ``generalized force'' acting on the spacetime via the
equations of motion, and the dissipation in the metric must
eventually be related to the fluctuations of this generalized
force for the relation to qualify as an FDR.

{}From this reasoning, we see that the stress energy bi-tensor and
its vacuum expectation value known as the noise kernel, are the
new ingredients in backreaction considerations. But these are
exactly the centerpiece in stochastic gravity. Therefore the
correct framework to address semiclassical backreaction problems
is stochastic gravity theory, where fluctuations and dissipation
are the equally essential components. The noise kernel for
quantum fields in Minkowski and de Sitter spacetime has been
carried out by Martin, Roura and Verdaguer \cite{MV1,MV2,RV02a},
for thermal fields in black hole spacetime and scalar fields in
general spacetimes by Campos, Hu and Phillips
\cite{CamHu,PH1,PH2,PH3}. Earlier, for cosmological backreaction
problems Hu and Sinha \cite{HuSin} derived a generalized
expression relating dissipation (of anisotropy in a Bianchi Type
I universes) and fluctuations (measured by particle numbers
created in neighboring histories). This example shows that one
can understand the backreaction of particle creation as a
manifestation of a (generalized) FDR.

As an illustration of the application of stochastic gravity theory
we outline the steps in a black hole backreaction calculation,
focusing on the manageable quasi-static class. We adopt the
Hartle-Hawking picture \cite{HarHaw76} where the black hole is
bathed eternally -- actually in quasi-thermal equilibrium -- in
the Hawking radiance it emits. It is described here by a massless
scalar quantum field at the Hawking temperature. As is
well-known, this quasi-equilibrium condition is possible only if
the black hole is enclosed in a box of size suitably larger than
the event horizon.
We can divide our consideration into the far field case and the
near horizon case. Campos and Hu \cite{CamHu} have treated a
relativistic thermal plasma in a weak gravitational field.  Since
the far field limit of a Schwarzschild metric is just the
perturbed Minkowski spacetime, one can perform a perturbation
expansion off hot flat space using the thermal Green functions
\cite{GibPer} \footnote{Strictly speaking the location of the box
holding the black hole in equilibrium with its thermal radiation
is as far as one can go, thus the metric may not reach the
perturbed Minkowski form. But one can also put the black hole and
its radiation in an anti-de Sitter space \cite{HawPag}, which
contains such a region.}. Hot flat space has been studied before
for various purposes. See e.g., \cite{GPY82,Reb91,ABF94}. Campos
and Hu derived a stochastic CTP effective action and from it an
equation of motion, the Einstein Langevin equation, for the
dynamical effect of a scalar quantum field on a background
spacetime \footnote{To perform calculations leading to the
Einstein-Langevin equation one needs to begin with a
self-consistent solution of the semiclassical Einstein equation
for the thermal field and the perturbed background spacetime. For
a black hole background, a semiclassical gravity solution is
provided by York \cite{York}. For a Robertson-Walker background
with thermal fields it is given by Hu \cite{Hu81}.}. Recently
Sinha, Raval and Hu \cite{SRH} outlined a strategy for treating
the near horizon case, following the same scheme of Campos and
Hu. In both cases two new terms appear which are absent in SCG
considerations: a nonlocal dissipation and a (generally colored)
noise kernel. When one takes the noise average one recovers
York's \cite{York} semiclassical equations for radially perturbed
quasi-static black holes. For the near horizon case one cannot
obtain the full details yet, because the Green function for a
scalar field in the Schwarzschild metric comes only in an
approximate form (e.g. Page approximation \cite{Page82}), which,
though reasonably accurate for the stress tensor, fails poorly
for the noise kernel \cite{PH2,PH3}. In addition a formula is
derived in \cite{SRH} expressing the CTP effective action in
terms of the Bogolyubov coefficients. Since it measures not only
the number of particles created, but also the difference of
particle creation in alternative histories, this provides a useful
avenue to explore the wider set of issues in black hole physics
related to noise and fluctuations.




Since backreaction calculations in semiclassical gravity has been
under study for a much longer time than in stochastic gravity we
will concentrate on explaining how the new stochastic features
arise from the framework of semiclassical gravity, i.e., noise and
fluctuations and their consequences. Technically the goal is to
obtain an influence action for this model of a black hole coupled
to a scalar field and to derive an Einstein-Langevin equation
from it. As a by-product, from the fluctuation-dissipation
relation, one can derive the vacuum susceptibility function and
the isothermal compressibility function for black holes, two
quantities of fundamental interest in characterizing the
nonequilibrium thermodynamic properties of black holes.

\subsection{The Model}

In this model the black hole spacetime is described by a
spherically symmetric static metric with line element of the
following general form written in advanced time
Eddington-Finkelstein coordinates \be ds^2 =
g_{\mu\nu}dx^{\mu}dx^{\nu} = -e^{2\psi}\left(1 - {2m\over
r}\right)dv^2 + 2 e^{2\psi}dvdr + r^2~d{\Omega}^2
\label{ssmetric} \te where $\psi = \psi(r)$ and $m = m(r)$ , $ v
= t + r + 2Mln\left({r\over 2M} -1 \right)$ and $d{\Omega}^2$ is
the line element on the two sphere. Hawking radiation is
described by a massless, conformally coupled quantum scalar field
$\phi$ with the classical action \be S_m[\phi, g_{\mu\nu}] =
-{\ha}\int d^n x \sqrt{-g}[g^{\mu\nu}\partial_{\mu}\phi
\partial_{\nu}\phi + \xi(n) R{\phi}^2] \label{phiact} \te where
$\xi(n) = {(n-2)\over 4(n-1)}$ ($n$ is the dimension of
spacetime) and $R$ is the curvature scalar of the spacetime it
lives in.

Let us consider linear perturbations $h_{\mu\nu}$ off a
background Schwarzschild metric $g^{(0)}_{\mu\nu}$ \be g_{\mu\nu}
= g^{(0)}_{\mu\nu} + h_{\mu\nu} \label{linearize} \te with
standard line element \be (ds^2)^0 = \left( 1 - {2M\over
r}\right)dv^2 + 2dvdr + r^2d{\Omega}^2 \label{schwarz} \te We
look for this class of perturbed metrics in the form given by
(\ref{ssmetric}), (thus restricting our consideration only to
spherically symmetric perturbations): \be e^\psi \simeq  1+
\epsilon \rho(r) \label{rho} \te and \be m \simeq M[ 1 + \epsilon
\mu (r)] \label{mu} \te where ${\epsilon\over \lambda M^2} =
{1\over 3}a T_H^4 ;$ $ a ={{\pi}^2\over 30} ; \lambda =
90(8^4)\pi^2$. $T_H$ is the Hawking temperature. This particular
parametrization of the perturbation is chosen following York's
\cite{York} notation. Thus the only non-zero components of
$h_{\mu\nu}$ are \be h_{vv} = -\left((1 - {2M\over r})2\epsilon
\rho(r) + {2M\epsilon \mu (r)\over r}\right) \label{hvv} \te and
\be h_{vr} = \epsilon\rho (r) \label{hvr} \te So this represents
a metric with small static and radial perturbations about a
Schwarzschild black hole. The initial quantum state of the scalar
field is taken to be the Hartle Hawking vacuum, which is
essentially a thermal state at the Hawking temperature and it
represents a black hole in (unstable) thermal equilibrium with
its own Hawking radiation. In the far field limit, the
gravitational field is described by a linear perturbation from
Minkowski spacetime. In equilibrium  the thermal bath can be
characterized by a relativistic fluid with a four-velocity
(time-like normalized vector field) $u^\mu$, and temperature in
its own rest frame $\beta^{-1}$.

To facilitate later comparisons with our program we briefly
recall York's work \cite{York}.\footnote{See also work by
Hochberg and Kephart \cite{HK} for a massless vector field,
Hochberg, Kephart and York \cite{HKY} for a massless spinor
field, and Anderson, Hiscock, Whitesell, and York \cite{AHWY} for
a quantized massless scalar field with arbitrary coupling to
spacetime curvature} He considered the semiclassical Einstein
equation \be G_{\m\n} (g_{\alpha \beta}) = \k \langle
T_{\m\n}\rangle \te with $G_{\mu\nu} \simeq G^{(0)}_{\mu\nu} +
\delta G_{\mu\nu}$ where $G^{(0)}_{\mu\nu}$ is the Einstein
tensor for the background spacetime. The zeroth order solution
gives a background metric in empty space, i.e, the Schwarzschild
metric. $\delta G_{\mu\nu}$ is the linear correction to the
Einstein tensor in the perturbed metric. The semiclassical
Einstein equation  in this approximation therefore reduces to \be
\delta G_{\mu\nu}(g^{(0)}, h) = \kappa \langle T_{\mu\nu} \rangle
\label{pertscee} \te York solved this equation to first order by
using the expectation value of the energy momentum tensor for a
conformally coupled scalar field in the Hartle-Hawking vacuum in
the unperturbed (Schwarzschild) spacetime on the right hand side
and using (\ref{hvv}) and (\ref{hvr}) to calculate $\delta
G_{\mu\nu}$ on the left hand side. Unfortunately, no exact
analytical expression is available for the $\langle
T_{\mu\nu}\rangle$ in a Schwarzschild metric with the quantum
field in the Hartle-Hawking vacuum that goes on the right hand
side. York therefore uses the approximate expression given by Page
\cite{Page82} which is known to give excellent agreement with
numerical results. Page's approximate expression for $\langle
T_{\mu\nu}\rangle$ was constructed using a thermal Feynman Green's
function obtained  by a conformal transformation of a WKB
approximated Green's function for an optical Schwarzschild
metric. York then solves the semiclassical Einstein equation
(\ref{pertscee}) self consistently to obtain the corrections to
the background metric induced by the backreaction encoded in the
functions $\mu(r)$ and $\rho(r)$. There was no mention of
fluctuations or its effects. As we shall see, in the language of
the previous section, the semiclassical gravity procedure which
York followed working at the equation of motion level is
equivalent to looking at the noise-averaged backreaction effects.

\subsection{CTP Effective Action for the Black Hole}


We first derive the CTP effective action for the model described
in the previous section. Using the  metric (\ref{schwarz}) (and
neglecting the surface terms that appear in an integration by
parts) we have  the action for the scalar field  written
perturbatively as
\begin{equation}
   S_m[\phi,h_{\mu\nu}]
        \ = \  {1\over 2}\int d^nx{\sqrt{-g^{(0)}}}\ \phi
               \left[ \Box^{(0)} + V^{(1)} + V^{(2)} + \cdots
              \right] \phi,
\label{phipert}
\end{equation}
where the first and second order perturbative operators $V^{(1)}$
and $V^{(2)}$ are given by
\begin{eqnarray}
V^{(1)}   & \equiv  & - {1\over \sqrt{-g^{(0)}}} \left\{
[\partial_\mu\left(\sqrt{-g^{(0)}}\bar h^{\mu\nu}(x)\right)]
                                \partial_\nu
                              +\bar h^{\mu\nu}(x)\partial_\mu
                              \partial_\nu
                            +\xi(n) R^{(1)}(x)
                     \right\},
               \nonumber \\
V^{(2)}
    &  \equiv & - {1\over \sqrt{-g^{(0)}}}
\left\{ [\partial_\mu \left(\sqrt{-g^{(0)}} \hat
h^{\mu\nu}(x)\right)]
                              \partial_\nu
                            +\hat h^{\mu\nu}(x)\partial_\mu
                            \partial_\nu
                          -\xi(n) ( R^{(2)}(x)
                               +{1\over 2}h(x)R^{(1)}(x))\right\}.
\end{eqnarray}
In the above expressions, $R^{(k)}$ is the $k$-order term in the
pertubation $h_{\mu\nu}(x)$ of the scalar curvature $R$ and $\bar
h_{\mu\nu}$ and $\hat h_{\mu\nu}$ denote a linear and a quadratic
combination of the perturbation, respectively,
\begin{eqnarray}
   \bar h_{\mu\nu}
        &  \equiv  & h_{\mu\nu} - {1\over 2} h g^{(0)}_{\mu\nu},
                     \nonumber \\
   \hat h_{\mu\nu}
        &  \equiv  & h^{\,\, \alpha}_\mu h_{\alpha\nu}
                      -{1\over 2} h h_{\mu\nu}
                      +{1\over 8} h^2 g^{(0)}_{\mu\nu}
                      -{1\over 4} h_{\alpha\beta}h^{\alpha\beta} g^{(0)}_{\mu\nu}.
   \label{eq:def bar h}
\end{eqnarray}
{}From quantum field theory in curved spacetime considerations
discussed above we take the following action for the gravitational
field:
\begin{eqnarray}
   S_g[g_{\mu\nu}]
        & \ = \ & {1\over {(16 \pi G)^{\frac{n-2}{2}}}}\int d^nx\ \sqrt{-g(x)}R(x)
                +{\alpha\bar\mu^{n-4}\over4(n-4)}
                   \int d^nx\ \sqrt{-g(x)} \nonumber \\
        &       &
                   \left\{ 3R_{\mu\nu\alpha\beta}(x)
                           R^{\mu\nu\alpha\beta}(x)
                         -\left[ 1-360 \left(\xi(n) - {1\over6}\right)^2
                          \right]R^2(x)
                   \right\}.
\end{eqnarray}
The first term is the classical Einstein-Hilbert action and the
second term is the counterterm in four dimensions used  to
renormalize the divergent effective action. In this action
$\ell^2_P = 16\pi G_N$, $\alpha = (2880\pi^2)^{-1}$ and $\bar\mu$
is an arbitrary mass scale.

We are interested in computing the CTP effective action
(\ref{phipert}) for the matter action and when the field $\phi$ is
initially in the Hartle-Hawking vacuum. This is equivalent to
saying that the initial state of the field is described by a
thermal density matrix at a finite temperature $T = T_H$. The CTP
effective action at finite temperature $T \equiv 1/\beta$ for
this model is given by (for details see \cite{CamHu})
\begin{equation}
   S_{rm eff}^\beta [h^\pm_{\mu\nu}]
        \ = \ S_g[h^+_{\mu\nu}]
             -S_g[h^-_{\mu\nu}]
             -{i\over2}Tr\{ \ln\bar G^\beta_{ab}[h^\pm_{\mu\nu}]\},
   \label{eq:eff act two fields}
\end{equation}
where $\pm$ denote the forward and backward time path of the CTP
formalism and $\bar G^\beta_{ab}[h^\pm_{\mu\nu}]$ is the complete
$2\times 2$ matrix propagator ($a$ and $b$ take $\pm$ values:
$G_{++},G_{+-}$ and $G_{--}$ correspond to the Feynman, Wightman
and Schwinger Greens functions respectively) with thermal boundary
conditions for the differential operator $\sqrt{-g^{(0)}}(\Box +
V^{(1)} + V^{(2)} + \cdots)$. 
The actual form of $\bar G^\beta_{ab}$ cannot be explicitly
given. However, it is easy to obtain a perturbative expansion in
terms of $V^{(k)}_{ab}$, the $k$-order matrix version of the
complete differential operator defined by $V^{(k)}_{\pm\pm}
\equiv \pm V^{(k)}_{\pm}$ and $V^{(k)}_{\pm\mp} \equiv 0$, and
$G^\beta_{ab}$, the thermal matrix propagator for a massless
scalar field in Schwarzschild spacetime . To second order $\bar
G^\beta_{ab}$ reads,
\begin{eqnarray}
   \bar G^\beta_{ab}
        \ = \  G^\beta_{ab}
              -G^\beta_{ac}V^{(1)}_{cd}G^\beta_{db}
              -G^\beta_{ac}V^{(2)}_{cd}G^\beta_{db}
              +G^\beta_{ac}V^{(1)}_{cd}G^\beta_{de}
               V^{(1)}_{ef}G^\beta_{fb}
              +\cdots
\end{eqnarray}
Expanding the logarithm and dropping one term independent of the
perturbation $h^\pm_{\mu\nu}(x)$, the CTP effective action may be
perturbatively written as,
\begin{eqnarray}
   S_{\rm eff}^\beta [h^\pm_{\mu\nu}]
        & = &  S_g[h^+_{\mu\nu}] - S_g[h^-_{\mu\nu}]
                \nonumber \\
        &       & +{i\over2}Tr[ V^{(1)}_{+}G^\beta_{++}
                               -V^{(1)}_{-}G^\beta_{--}
                               +V^{(2)}_{+}G^\beta_{++}
                               -V^{(2)}_{-}G^\beta_{--}
                              ]
                \nonumber \\
        &       & -{i\over4}Tr[  V^{(1)}_{+}G^\beta_{++}
                                 V^{(1)}_{+}G^\beta_{++}
                               + V^{(1)}_{-}G^\beta_{--}
                                 V^{(1)}_{-}G^\beta_{--}
                               -2V^{(1)}_{+}G^\beta_{+-}
                                 V^{(1)}_{-}G^\beta_{-+}
                              ].
   \label{eq:effective action}
\end{eqnarray}
In computing the traces, some terms containing divergences are
canceled using counterterms introduced in the classical
gravitational action after dimensional regularization.

\subsection{Near Flat Case}

At this point we divide our considerations into two cases. In the
far field limit  $h_{\mu\nu}$ represent perturbations about flat
space, i.e., $g^{(0)}_{\mu\nu}= \eta_{\m\n}$. The exact
``unperturbed" thermal propagators for scalar fields are known,
i.e., the Euclidean propagator with periodicity $\beta$. Using
the Fourier transformed  form (those quantities are denoted with a
tilde) of the thermal propogators $\tilde G^\beta_{ab}(k)$ , the
trace terms of the form
$Tr[V^{(1)}_{a}G^\beta_{mn}V^{(1)}_{b}G^\beta_{rs}]$ can be
written as \cite{CamHu},
\begin{equation}
   Tr[V^{(1)}_{a}G^\beta_{mn}V^{(1)}_{b}G^\beta_{rs}]
        \ = \  \int d^nxd^nx'\
               h^a_{\mu\nu}(x)h^b_{\alpha\beta}(x')
               \int {d^nk\over(2\pi)^n}{d^nq\over(2\pi)^n}
               e^{ik\cdot (x-x')}
               \tilde G^\beta_{mn}(k+q)\tilde G^\beta_{rs}(q)
               \kl{T}{}{q,k},
   \label{eq:trace}
\end{equation}
where the tensor $\kl{T}{}{q,k}$ is defined in \cite{CamHu} after
an  expansion in terms of a basis of 14 tensors \cite{Reb91}. In
particular, the last trace of (\ref{eq:effective action}) may be
split in two different kernels $\kl{N}{}{x-x'}$ and
$\kl{D}{}{x-x'}$,
\begin{equation}
   {i\over2}Tr[V^{(1)}_{+}G^\beta_{+-}V^{(1)}_{-}G^\beta_{-+}]
        \ = \ -\int d^4xd^4x'\
               h^+_{\mu\nu}(x)h^-_{\alpha\beta}(x')
               [   \kl{D}{}{x-x'}
                +i \kl{N}{}{x-x'}
               ].
\end{equation}
One can express the Fourier transforms of these kernels,
respectively, as
\begin{eqnarray}
   \kl{\tilde N}{}{k}
        & \ = \ & \pi^2\int {d^4q\over(2\pi)^4}\
                  \left\{ \theta(k^o+q^o)\theta(-q^o)
                         +\theta(-k^o-q^o)\theta(q^o)
                         +n_\beta(|q^o|)+n_\beta(|k^o+q^o|)
                  \right.
                \nonumber \\
        &       & \hskip2cm
                  \left. +2n_\beta(|q^o|)n_\beta(|k^o+q^o|)
                  \right\}\delta(q^2)\delta[(k+q)^2]\kl{T}{}{q,k},
   \label{eq:N}
\end{eqnarray}
\begin{eqnarray}
   \kl{\tilde D}{}{k}
        & \ = \ & -i\pi^2\int {d^4q\over(2\pi)^4}\
                  \left\{ \theta(k^o+q^o)\theta(-q^o)
                         -\theta(-k^o-q^o)\theta(q^o)
                         +sg(k^o+q^o) n_\beta(|q^o|)
                  \right.
                \nonumber \\
        &       & \hskip2cm
                  \left. -sg(q^o)n_\beta(|k^o+q^o|)
                  \right\}\delta(q^2)\delta[(k+q)^2]\kl{T}{}{q,k}.
   \label{eq:D}
\end{eqnarray}
Using the property $\kl{T}{}{q,k} = \kl{T}{}{-q,-k}$, it is easy
to see that $\kl{N}{}{x-x'}$ is symmetric and $\kl{D}{}{x-x'}$
antisymmetric in their arguments; that is, $\kl{N}{}{x} =
\kl{N}{}{-x}$ and $\kl{D}{}{x} = -\kl{D}{}{-x}$.

The physical meanings of these kernels can be extracted if we
write the renormalized CTP effective action at finite temperature
(\ref{eq:effective action}) in an influence functional form
\cite{ifqbm}. N, the imaginary part of the CTP effective action
can be identified with the noise kernel and D, the antisymmetric
piece of the real part, with the dissipation kernel. Campos and
Hu \cite{CamHu} have shown that these kernels identified as such
indeed satisfy a thermal FDR.

If we denote the difference and the sum of the perturbations
$h^\pm_{\mu\nu}$ defined along each branch $C_\pm$ of the complex
time path of integration $C$ by $[h_{\mu\nu}] \equiv h^+_{\mu\nu}
- h^-_{\mu\nu}$ and $\{h_{\mu\nu}\} \equiv h^+_{\mu\nu} +
h^-_{\mu\nu}$, respectively, the influence functional form of the
thermal CTP effective action may be written to second order in
$h_{\mu\nu}$ as,
\begin{eqnarray}
   S_{\rm eff}^\beta [h^\pm_{\mu\nu}]
        & \ \simeq \ & {1\over 2(16 \pi G_N)} \int d^4x\ d^4x'\
                       [h_{\mu\nu}](x)\kl{L}{(o)}{x-x'}
                       \{h_{\alpha\beta}\}(x')
                     \nonumber \\
        &            &+{1\over2}\int d^4x\
                       [h_{\mu\nu}](x)T^{\mu\nu}_{(\beta)}
                     \nonumber \\
        &            &+{1\over2}\int d^4x\ d^4x'\
                       [h_{\mu\nu}](x)\kl{H}{}{x-x'}
                       \{h_{\alpha\beta}\}(x')
                     \nonumber \\
        &            &-{1\over2}\int d^4x\ d^4x'\
                       [h_{\mu\nu}](x)\kl{D}{}{x-x'}
                       \{h_{\alpha\beta}\}(x')
                     \nonumber \\
        &            &+{i\over2}\int d^4x\ d^4x'\
                       [h_{\mu\nu}](x)\kl{N}{}{x-x'}
                       [h_{\alpha\beta}](x').
\label{CTPbh}
\end{eqnarray}
The first line is the Einstein-Hilbert action to second order in
the perturbation $h^\pm_{\mu\nu}(x)$. $\kl{L}{(o)}{x}$ is a
symmetric kernel ({\sl i.e.} $\kl{L}{(o)}{x}$ =
$\kl{L}{(o)}{-x}$). In the near flat case its Fourier transform
is given by
\begin{equation}
   \kl{\tilde L}{(o)}{k}
        \ = \ {1\over4}\left[ - k^2 \kl{T}{1}{q,k}
                              +2k^2 \kl{T}{4}{q,k}
                              + \kl{T}{8}{q,k}
                              -2\kl{T}{13}{q,k}
                       \right].
\end{equation}
The fourteen elements of the tensor basis $\kl{T}{i}{q,k}$
($i=1,\cdots,14$) are defined in \cite{Reb91}. The second is a
local term linear in $h^\pm_{\mu\nu}(x)$.  Only when far away
from the hole that it takes the form of the stress tensor of
massless scalar particles at temperature $\beta^{-1}$, which has
the form of a perfect fluid stress-energy tensor
\begin{equation}
   T^{\mu\nu}_{(\beta)}
        \ = \ {\pi^2\over30\beta^4}
              \left[ u^\mu u^\nu + {1\over3}(\eta^{\mu\nu}+u^\mu u^\nu)
              \right],
\end{equation}
where $u^\mu$ is the four-velocity of the plasma \footnote{In the
far field limit, taking into account the four-velocity $u^\mu$ of
the fluid, a manifestly Lorentz-covariant approach to thermal
field theory may be used \cite{Wel82}. However, in order to
simplify the involved tensorial structure we work in the
co-moving coordinate system of the fluid where $u^\mu =
(1,0,0,0)$.} and the factor ${\pi^2\over30\beta^4}$ is the
familiar thermal energy density for massless scalar particles at
temperature $\beta^{-1}$. In the third line, the Fourier
transform of the symmetric kernel $\kl{H}{}{x}$ can be expressed
as
\begin{eqnarray}
   \kl{\tilde H}{}{k}
        & \ = \ &  -{\alpha k^4\over4}
                   \left\{ {1\over2}\ln {|k^2|\over\mu^2}\kl{Q}{}{k}
                          +{1\over3}\kl{\bar Q}{}{k}
                   \right\}
                \nonumber \\
        &       &  +{\pi^2\over180\beta^4}
                   \left\{ - \kl{T}{1}{u,k}
                           -2\kl{T}{2}{u,k}
                           + \kl{T}{4}{u,k}
                           +2\kl{T}{5}{u,k}
                   \right\}
                \nonumber \\
        &       &  +{\xi\over96\beta^2}
                   \left\{    k^2 \kl{T}{1}{u,k}
                           -2 k^2 \kl{T}{4}{u,k}
                           -      \kl{T}{8}{u,k}
                           +2     \kl{T}{13}{u,k}
                   \right\}
                \nonumber \\
        &       &  +\pi\int {d^4q\over(2\pi)^4}\
                   \left\{ \delta(q^2)n_\beta(|q^o|)
                           {\cal P}\left[ {1\over(k+q)^2}
                                   \right]
                          +\delta[(k+q)^2]n_\beta(|k^o+q^o|)
                           {\cal P}\left[ {1\over q^2}
                                   \right]
                   \right\}\kl{T}{}{q,k},
   \label{eq:grav pol tensor}
\end{eqnarray}
where $\mu$ is a simple redefinition of the renormalization
parameter $\bar\mu$ given by $\mu \equiv \bar\mu \exp
({23\over15} + {1\over2}\ln 4\pi - {1\over2}\gamma)$, and the
tensors $\kl{Q}{}{k}$ and $\kl{\bar Q}{}{k}$ are defined,
respectively, by
\begin{eqnarray}
   \kl{Q}{}{k}
        & \ = \ & {3\over2} \left\{               \kl{T}{1}{q,k}
                                    -{1\over k^2} \kl{T}{8}{q,k}
                                    +{2\over k^4} \kl{T}{12}{q,k}
                            \right\}
                \nonumber \\
        &       &-[1-360(\xi-{1\over6})^2]
                  \left\{               \kl{T}{4}{q,k}
                          +{1\over k^4} \kl{T}{12}{q,k}
                          -{1\over k^2} \kl{T}{13}{q,k}
                  \right\},
   \label{eq:Q tensor}
\end{eqnarray}
\begin{equation}
   \kl{\bar Q}{}{k}
        \ = \  [1+576(\xi-{1\over6})^2-60(\xi-{1\over6})(1-36\xi')]
                  \left\{               \kl{T}{4}{q,k}
                          +{1\over k^4} \kl{T}{12}{q,k}
                          -{1\over k^2} \kl{T}{13}{q,k}
                  \right\}.
\end{equation}
In the above and subsequent equations, we denote the coupling
parameter in four dimensions $\xi(4)$ by $\xi$ and consequently
$\xi'$ means $d\xi(n)/dn$ evaluated at $n=4$. $\kl{\tilde
H}{}{k}$ is the complete contribution of a free massless quantum
scalar field to the thermal graviton polarization
tensor\cite{Reb91,ABF94} and it is responsible for the
instabilities found in flat spacetime at finite temperature
\cite{GPY82,Reb91,ABF94} \footnote{Note that the addition of the
contribution of other kinds of matter fields to the effective
action, even graviton contributions, does not change the tensor
structure of these kernels and only the overall factors are
different to leading order \cite{Reb91}.}. Eq.~(\ref{eq:grav pol
tensor}) reflects the fact that the kernel $\kl{\tilde H}{}{k}$
has thermal as well as non-thermal contributions. Note that it
reduces to the first term in the zero temperature limit
($\beta\rightarrow\infty$)
\begin{equation}
   \kl{\tilde H}{}{k}
        \ \simeq \ -{\alpha k^4\over4}
                     \left\{ {1\over2}\ln {|k^2|\over\mu^2}\kl{Q}{}{k}
                            +{1\over3}\kl{\bar Q}{}{k}
                     \right\}.
\end{equation}
and at high temperatures the leading term ($\beta^{-4}$) may be
written as
\begin{equation}
   \kl{\tilde H}{}{k}
        \ \simeq \ {\pi^2\over30\beta^4}
                    \sum^{14}_{i=1}
                    \mbox{\rm H}_i(r) \kl{T}{i}{u,K},
\end{equation}
where we have introduced the dimensionless external momentum
$K^\mu \equiv k^\mu/|\vec{k}| \equiv (r,\hat k)$. The $\mbox{\rm
H}_i(r)$ coefficients were first given in \cite{Reb91} and
generalized to the next-to-leading order ($\beta^{-2}$) in
\cite{ABF94}. (They are given with the MTW sign convention  in
\cite{CamHu}.)

Finally, as defined above, $\kl{N}{}{x}$ is the noise kernel
representing the random fluctuations of the thermal radiance and
$\kl{D}{}{x}$ is the dissipation kernel, describing the
dissipation of energy of the gravitational field.


\subsection{Near Horizon Case}

In this case, since the perturbation is taken around the
Schwarzschild spacetime, exact expressions for the corresponding
unperturbed propagators $G^\beta_{ab}[h^\pm_{\mu\nu}]$ are not
known. Therefore apart from the approximation of computing the CTP
effective action to certain order in perturbation theory, an
appropriate approximation scheme for the unperturbed Green's
functions is also required. This feature manifested itself in
York's calculation of backreaction as well, where, in writing the
$\langle T_{\mu \nu}\rangle$ on the right hand side of the
semiclassical Einstein equation in the unperturbed Schwarzschild
metric, he had to use an approximate expression for $\langle
T_{\mu\nu} \rangle$ in the Schwarzschild metric given by Page
\cite{Page82}. The additional complication here is that while to
obtain $\langle T_{\mu\nu}\rangle$ as in York's calculation, the
knowledge of only the thermal Feynman Green's function is
required, to calculate the CTP effective action one needs the
knowledge of the full matrix propagator, which involves the
Feynman, Schwinger and Wightman functions.

It is indeed possible to construct the full thermal matrix
propagator $G^\beta_{ab}[h^\pm_{\mu\nu}]$ based on Page's
approximate Feynman Green's function by using identities relating
the Feynman Green's function with the other Green's functions
with different boundary conditions. One can then proceed to
explicitly compute a CTP effective action and hence the influence
functional based on this approximation. However, we desist from
delving into such a calculation for the following reason. Our
main interest in performing such a calculation is to identify and
analyze the noise term which is the new ingredient in the
backreaction. We have mentioned that the noise term gives a
stochastic contribution $\xi^{\mu\nu}$ to the Einstein-Langevin
equation (\ref{2.11}). We had also stated that this term is
related to the variance of fluctuations in $T_{\mu\nu}$, i.e,
schematically, to $\langle T^2_{\mu\nu}\rangle$. However, a
calculation of $\langle T^2_{\mu\nu}\rangle$ in the
Hartle-Hawking state in a Schwarzschild background using the Page
approximation was performed by Phillips and Hu \cite{PH1,PH2,PH3}
and it was shown that though the approximation is excellent as
far as $\langle T_{\mu\nu}\rangle$ is concerned, it gives
unacceptably large errors for $\langle T^2_{\mu\nu}\rangle$ at
the horizon.
In fact, similar errors will be propagated in the non-local
dissipation term as well, because both terms originate from the
same source, that is, they come from the last trace term in
(\ref{eq:effective action}) which contains terms quadratic in the
Green's function. However, the Influence Functional or CTP
formalism itself does not depend on the nature of the
approximation, so we will attempt to exhibit the general
structure of the calculation without resorting to a specific form
for the Greens function and conjecture on what is to be expected.
A more accurate computation can be performed using this formal
structure once a better approximation becomes available.

The general structure of the CTP effective action arising from
the calculation of the traces in equation (\ref{eq:effective
action}) remains the same. But to write down explicit expressions
for the non-local kernels one requires the input of the explicit
form of $G^\beta_{ab}[h^\pm_{\mu\nu}]$  in the Schwarzschild
metric, which is not available in closed form. We can make some
general observations about the terms in there. The first line
containing L does not have an explicit Fourier representation as
given in the far field case, neither will $T_{(\beta)}^{\mu\nu}$
in the second line representing the zeroth order contribution to
$\langle T_{\mu\nu} \rangle$ have a perfect fluid form. The third
and fourth terms containing the remaining quadratic component of
the real part of the effective action will not have any simple or
even complicated analytic form. The symmetry properties of the
kernels $H^{\mu\nu,\alpha\beta}(x,x')$ and
$D^{\mu\nu,\alpha\beta}(x,x')$ remain intact, i.e., they are
respectively even and odd in $x,x'$. The last term in the CTP
effective action gives the imaginary part of the effective action
and the kernel $N(x,x')$ is symmetric.

Continuing our general observations from this CTP effective
action, using the connection between this thermal CTP effective
action to the influence functional \cite{Su,CH94} via an equation
in the schematic form  (\ref{ctpif}).
We see that the nonlocal imaginary term containing the kernel
$N^{\mu\nu,\alpha\beta}(x,x')$ is responsible for the generation
of the stochastic noise term in the Einstein-Langevin equation
and the real non-local term containing kernel
$D^{\mu\nu,\alpha\beta}(x,x')$ is responsible for the non-local
dissipation term. To derive the Einstein-Langevin equation we
first construct the stochastic effective action (\ref{stochastic
eff action}).
We then derive the equation of motion, as shown earlier in (\ref{eq of motion}), 
by taking its functional derivative with respect to
$[h_{\mu\nu}]$ and equating it to zero.
With the identification of noise and dissipation kernels, one can
write down a linear, non-local relation of the form, \be N(t-t') =
~\int~d(s -s')K(t-t',s-s')\gamma(s -s') \label{FDR}, \te where
$D(t,t')=-\partial_{t'}\gamma (t,t')$.  This is the general
functional form of a Fluctuation-Dissipation relation (FDR) and
$K(t,s)$ is called the fluctuation-dissipation kernel
\cite{ifqbm}. In the present context this relation depicts the
backreaction of thermal Hawking radiance for a black hole in
quasi-equilibrium.


\subsection{Einstein-Langevin equation}

In this section we show how  a semiclassical Einstein-Langevin
equation  can be derived from the previous thermal CTP effective
action. This equation depicts the stochastic evolution of the
perturbations of the black hole under the influence of the
fluctuations of the thermal scalar field.

The influence functional ${\cal F}_{\rm IF} \equiv \exp (iS_{\rm
IF})$ previously introduced in Eq. (\ref{influence functional})
can be written in terms of the the CTP effective action $S_{\rm
eff} ^\beta [h^\pm_{\mu\nu}]$ derived in equation (\ref{CTPbh})
using Eq.(\ref{ctpif}).  The Einstein-Langevin equation follows
from taking the functional derivative of the stochastic effective
action (\ref{stochastic eff action}) with respect to
$[h_{\mu\nu}](x)$ and imposing $[h_{\mu\nu}](x) = 0$
This leads to
\begin{equation}
   {1\over\ell^2_P}
   \int d^4x'\ \kl{L}{(o)}{x-x'} h_{\alpha\beta}(x')
  +{1\over2}\ T^{\mu\nu}_{(\beta)}
  +\int d^4x'\ \left( \kl{H}{}{x-x'}
                     -\kl{D}{}{x-x'}
               \right) h_{\alpha\beta}(x')
  +\xi^{\mu\nu}(x)
        \ = \ 0.
\end{equation}
where
\begin{equation}
   \langle \xi^{\mu\nu}(x) \xi^{\alpha\beta}(x') \rangle_j
        \ = \ \kl{N}{}{x-x'},
   \label{eq:correlation}
\end{equation}
In the far field limit this equation should reduce to that
obtained by Campos and Hu \cite{CamHu}: For gravitational
perturbations $h^{\mu\nu}$ defined in (\ref{eq:def bar h}) under
the harmonic gauge $\bar h^{\mu\nu}_{\,\,\,\,\, ,\nu} = 0$, their
Einstein-Langevin equation is given by
\begin{equation}
   \Box\bar h^{\mu\nu}(x)
         + {1 \over (16 \pi G_N^2)}
               \left\{ T^{\mu\nu}_{(\beta)}
                      +2P_{\rho\sigma,\alpha\beta}
                       \int d^4x'\ \left( \kl{H}{}{x-x'}
                                         -\kl{D}{}{x-x'}
                                   \right)\bar h^{\rho\sigma}(x')
                      +2\xi^{\mu\nu}(x)
               \right\} = 0,
\end{equation}
where the tensor $P_{\rho\sigma,\alpha\beta}$ is given by
\begin{equation}
   P_{\rho\sigma,\alpha\beta}
        \ = \ {1\over2}\left( \eta_{\rho\alpha}\eta_{\sigma\beta}
                             +\eta_{\rho\beta}\eta_{\sigma\alpha}
                             -\eta_{\rho\sigma}\eta_{\alpha\beta}
                       \right).
\end{equation}
The expression for  $P_{\rho\sigma,\alpha\beta}$ in the near
horizon limit of course cannot be expressed in such a simple form.
Note that this differential stochastic equation includes a
non-local term responsible for the dissipation of the
gravitational field and a noise source term which accounts for
the fluctuations of the quantum field . Note also that this
equation in combination with the correlation for the stochastic
variable (\ref{eq:correlation}) determine the two-point
correlation for the stochastic metric fluctuations $\langle \bar
h_{\mu\nu}(x) \bar h_{\alpha\beta}(x') \rangle_\xi$
self-consistently.

As we have seen before and here, the Einstein-Langevin equation is
a dynamical equation governing the dissipative evolution of the
gravitational field under the influence of the fluctuations of
the quantum field, which, in the case of black holes, takes the
form of thermal radiance. From its form we can see that even for
the quasi-static case under study the back reaction of Hawking
radiation on the black hole spacetime has an innate dynamical
nature.

For the far field case making use of the explict forms available
for the noise and dissipation kernels Campos and Hu \cite{CamHu}
formally proved the existence of a Fluctuation-Dissipation
Relation (FDR) at all temperatures between the quantum
fluctuations of the thermal radiance and the dissipation of the
gravitational field.  They also showed the formal equivalence of
this method with Linear Response Theory (LRT) for lowest order
perturbance of a near-equilibrium system, and how the response
functions such as the contribution of the quantum scalar field to
the thermal graviton polarization tensor can be derived. An
important quantity not usually obtained in LRT, but of equal
importance, manifest in the CTP stochastic approach is the noise
term arising from the quantum and statistical fluctuations in the
thermal field. The example given in this section shows that the
back reaction is intrinsically a dynamic process described (at
this level of sophistication) by the Einstein-Langevin equation.
By comparison, traditional LRT calculations cannot capture the
dynamics as fully and thus cannot provide a complete description
of the backreaction problem.


\subsection{Discussions}

We now draw some connection with related work. As remarked
earlier, except for the near-flat case, an analytic form of the
Green function is not available. Even the Page approximation
\cite{Page82} which gives unexpectedly good results for the stress
energy tensor has been shown to fail in the fluctuations of the
energy density \cite{PH2,PH3}. Thus using such an approximation
for the noise kernel will give unreliable results for the
Einstein-Langevin equation.  If we confine ourselves to Page's
approximation and derive the
equation of motion
without the stochastic term, we expect to recover York's
semiclassical Einstein's equation if one retains only the zeroth
order contribution, i.e, the first two terms in the expression
for the CTP effective action in Eq. (\ref{CTPbh}). Thus, this
offers a new route to arrive at York's semiclassical Einstein's
equations. Not only is it a derivation of York's result from a
different point of view, but it also shows how his result arises
as an appropriate limit of a more complete framework, i.e, it
arises when one averages over the noise. Another point worth
noting is that our treatment  will also yield a non-local
dissipation term arising from the fourth term in equation
(\ref{CTPbh}) in the CTP effective action which is absent in
York's treatment. This difference arises primarily due to the
difference in the way backreaction is treated,  at the level of
iterative approximations on the equation of motion as in York,
versus the treatment at the effective action level as pursued
here. In York's treatment, the Einstein tensor is computed to
first order in perturbation theory , while $\langle
T_{\mu\nu}\rangle$ on the right hand side of the semiclassical
Einstein equation is replaced by the zeroth order term. In the
effective action treatment the full effective action is computed
to second order in perturbation, and hence includes the higher
order non-local terms.

The other important conceptual point that comes to light from this
approach is that related to the Fluctuation-Dissipation Relation.
In the quantum Brownian motion analog (e.g., \cite{ifqbm} and
references therein), the dissipation of the energy of the
Brownian particle as it approaches equilibrium and the
fluctuations at equilibrium are connected by the Fluctuation -
Dissipation relation
Here the backreaction of quantum fields on black holes also
consists of two forms -- dissipation and fluctuation or noise,
corresponding to the real and imaginary parts of the influence
functional as embodied in the dissipation and noise kernels. A
FDR relation has been shown to exist for the near flat case by
Campos and Hu \cite{CamHu} and we anticipate that it should also
exist between the noise and dissipation kernels for the general
case, as it is a categorical relation \cite{ifqbm,Banff}. Martin
and Verdaguer have also proved the existence of a FDR when the
semiclassical background is a stationary spacetime and the
quantum field is in thermal equilibrium. Their result was then
extended to a conformal field in a conformaly stationary
background \cite{MV1}. The existence of a FDR for the black hole
case has been discussed by some authors previously
\cite{CanSci,Mottola}. In \cite{Vishu}, Hu, Raval and Sinha have
described how this approach and results differ from those of
previous authors. The FDR reveals an interesting connection
between black holes interacting with quantum fields and
non-equilibrium statistical mechanics. Even in its restricted
quasi-static form, this relation will allow us to study
\textit{nonequilibrium} thermodynamic properties of the black
hole under the influence of stochastic fluctuations of the energy
momentum tensor dictated by the noise terms.

There are limitations of a technical nature in the specific
example invoked here. For one we have to confine ourselves to
small perturbations about a background metric. For another, as
mentioned above, there is no
reliable approximation to the Schwarzschild thermal Green's
function to explicitly compute the noise and dissipation kernels.
This limits our ability to present explicit analytical
expressions for these kernels. One can try to improve on Page's
approximation by retaining terms to higher order. A less
ambitious first step could be to confine attention to the
horizon  and using approximations that are restricted to near the
horizon and work out the Influence Functional in this
regime. 

Yet another technical limitation of the specific example is the
following. Though we have allowed for backreaction effects to
modify the initial state in the sense that the temperature of the
Hartle-Hawking state gets affected by the backreaction, we have
essentially confined our analysis to a Hartle-Hawking thermal
state of the field. This analysis does not directly extend to a
more general class of states, for example to the case where the
initial state of the field is in the Unruh vacuum. Thus we will
not be able to comment on issues of the stability of an
\textit{isolated} radiating black hole under the influence of
stochastic fluctuations.

\section{Further Developments}

In this review  we have given two routes to the establishment of
stochastic gravity theory with derivation of the influence
functional  and the Einstein-Langevin equation (ELE). We also
showed two examples of how backreaction problems can be treated
by the influence functional or the CTP effective action method
leading to an ELE describing the dissipative dynamics of
spacetime driven by the fluctuations of the quantum field. In
both of these examples we considered linear perturbations off a
background spacetime, where the Green function for the quantum
scalar field is readily available and the calculation of the CTP
effective action can be performed by a perturbative expansion.
This is true for the family of cosmological spacetimes, because
of its high symmetry, and for the far field limit of the
Schwarzschild spacetime. In the near horizon case, calculation is
handicapped because an analytic form of the Green function is not
available. These limitations are of a technical nature. However,
the formulation of the problem and the overall strategy of attack
are general enough to be applicable to a wide range of
fluctuations and backreaction problems. This is the context where
stochastic gravity theory was historically invented and where it
can best be utilized.


We mention a number of ongoing research related to the topics
discussed in this review. On the theory side, Roura and Verdaguer
\cite{RV02b} has recently showed how stochastic gravity can be
related to the large $N$ limit of quantum metric fluctuations.
Given $N$ free matter fields weakly interacting with the
gravitational field, Hartle and Horowitz \cite{Hartle-Horowitz81}
and Tomboulis \cite{Tomboulis77} have shown that semiclassical
gravity can be obtained as the leading order large $N$ limit
(while keeping $N$ times the gravitational coupling constant
fixed). It is of interest to find out where in this setting can
one place the fluctuations of the quantum fields and the metric
fluctuations they induce; specifically, whether the inclusion of
these sources will lead to an Einstein-Langevin equation
\cite{CH94,HM3,HuSin,CamVer96,LomMaz97}, as it was derived
historically in other ways as described in the first part of this
review. This is useful because it may provide another pathway or
angle in connecting semiclassical to quantum gravity (Using
interacting quantum fields as example, a related idea is the
kinetic approach to quantum gravity described in \cite{kinQG}).

\subsection{Metric Fluctuations and Structure Formation}

Starobinsky's stochastic inflation \cite{stoinf} is an
interesting paradigm for treating structure formation  in that the
classical long wavelength modes governed by a stochastic equation
is driven by a noise originating from the quantum fluctuations of
the high frequency modes. However, there are conceptual and
technical problems  (see, e.g., \cite{cgea,CH94,strfor}).
Specifically, how the long wavelength modes turn classical and
how the quantum fluctuations act like noise. These bear on the
issues of decoherence of the mean field and the quantum field
origin of noise, two  issues at the foundations of all theories
of structure formation based on quantum fluctuations. Stochastic
gravity is closest in spirit to this paradigm and is thus expected
to be the proper theoretical framework for addressing these
outstanding issues.

In another related problem, standard theories of structure
formation from quantum (inflaton) fields and their fluctuations
are based on the quantization of the linear perturbations of both
the metric and the inflaton field \cite{Mukhanov92}. A question
naturally arises for stochastic gravity as to whether one can
recover the same equations for the quantum correlation functions
with the correlation function of the metric fluctuations. Recently
Roura and Verdaguer \cite{RV02a} have given an expression of the
two point function of metric perturbations in terms of the
stochastic source from the products of the Hadamard functions of
the inflaton field in a quasi- de Sitter spacetime, which is
connected to the stress energy tensor fluctuations through the
noise kernel. Although the gravitational fluctuations are assumed
to be classical in stochastic gravity, at least in the linear
regime, their correlation functions predicted by the
Einstein-Langevin equation gives the correct symmetrized quantum
two point functions \cite{CRVopensys}. This actually simplifies
the conventional approach.  Another advantage of the stochastic
gravity approach is that it can also tackle gravitational
fluctuations in inflationary models which are not driven by an
inflaton field, but by vacuum polarization effects such as the
trace anomaly (e.g., in Starobinsky inflation
\cite{Starobinsky80,Vilenkin85,Hawking01}).

\subsection{Metric Fluctuations in Black Holes}

In addition to the work described above by Campos, Hu, Raval and
Sinha \cite{CamHu,Vishu,SRH} and earlier work quoted therein, we
mention also some recent work on black hole metric fluctuations
and their effect on Hawking radiation. For example, Casher et al
\cite{CEIMP} and Sorkin \cite{Sorkin} have concentrated on the
issue of fluctuations of the horizon induced by a fluctuating
metric. Casher et al \cite{CEIMP} considers the fluctuations of
the horizon induced by the ``atmosphere" of high angular momentum
particles near the horizon, while Sorkin \cite{Sorkin} calculates
fluctuations of the shape of the horizon induced by the quantum
field fluctuations under a Newtonian approximations. Both group
of authors come to the conclusion that horizon fluctuations
become large at scales much larger than the Planck scale (note
Ford and Svaiter \cite{ForSva} later presented results contrary
to this claim). However, though these works do deal with
backreaction, the fluctuations considered do not arise as an
explicit stochastic noise term as in our treatment. It may be
worthwhile exploring the horizon fluctuations induced by the
stochastic metric in our model and comparing the conclusions with
the above authors. Barrabes et al \cite{Barrabes}  have
considered the propagation of null rays and massless fields in a
black hole fluctuating geometry and have shown that the
stochastic nature of the metric leads to a modified dispersion
relation and helps to confront the trans-Planckian frequency
problem. However, in this case the stochastic noise is put in by
hand and does not naturally arise from coarse graining as in the
quantum open systems approach.  It also does not take
backreaction into account. It will be interesting to explore how
a stochastic black hole metric, arising as a solution to the
Einstein-Langevin equation, hence fully incorporating
backreaction, would affect the trans-Planckian problem.

Ford and his collaborators  \cite{ForSva,ForWu,WuFor99} have also
explored the issue of metric fluctuations in detail and in
particular have studied the fluctuations of the black hole
horizon induced by metric fluctuations. However, the fluctuations
they considered are in the context of a fixed background and do
not relate to the backreaction.

Another work originating from the same vein of stochastic gravity
but not complying with the backreaction spirit is that of Hu and
Shiokawa \cite{HuShio}, who study novel effects associated with
electromagnetic wave propagation in a Robertson-Walker universe
and the Schwarzschild spacetime with a small amount of given
metric stochasticity. For the Schwarzschild metric, they find
that time-independent randomness can decrease the total
luminosity of Hawking radiation due to multiple scattering of
waves outside the black hole and gives rise to event horizon
fluctuations and fluctuations in the Hawking temperature. The
stochasticity in a background metric in their work is assumed
rather than derived (from quantum field fluctuations, as in this
work) and so is not in the same spirit of backreaction. But it is
interesting to compare their results with that of backreaction,
so one can begin to get a sense of the different sources of
stochasticity and their weights (see, e.g., \cite{stogra} for a
list of possible sources of stochasticity.)

In a subsequent paper Shiokawa \cite{Shio} showed that the scalar
and spinor waves in a stochastic spacetime behave similarly to
the electrons in a disordered system. Viewing this as a quantum
transport problem, he expressed the conductance and its
fluctuations in terms of a nonlinear sigma model in the closed
time path formalism and showed that the conductance fluctuations
are universal, independent of the volume of the stochastic region
and the amount of stochasticity. This result can have significant
importance in characterizing the mesoscopic behavior of
spacetimes resting between the semiclassical and the quantum
regimes.

The stochastic approach to the study of black hole backreaction
thus has a very rich structure and opens up many new avenues of
inquiry. In particular it provides the proper platform and
framework to launch a new program of research into the {\it
nonequilibrium black hole thermodyamics}.

As illustrated in the cosmological backreaction example stochastic
gravity theory can be applied to quasi-dynamic or even fully
dynamic problems such as black hole collapse, technical
difficulty of finding reasonable analytic approximations of the
Green function or numerical evaluation of mode-sums
notwithstanding. Our discussion shows that stochastic gravity
based on open systems concepts and the close-time-path or
influence functional methods is the preferred framework for
backreaction problems of dynamical spacetimes interacting with
quantum fields and is specially suitable for treating spacetime
fluctuations and non-equilibrium conditions of
matter fields.\\

\noindent {\bf Acknowledgements} The materials presented here
originated from research work of BLH with Antonio Campos, Alpan
Raval and Sukanya Sinha,  and of EV with Antonio Campos and
Rosario Martin. We thank them as well as Daniel Arteaga, Andrew
Matacz, Albert Roura, Nicholas Phillips, Tom Shiokawa, and Yuhong
Zhang for fruitful collaboration and their cordial friendship
since their Ph. D. days. We enjoy lively discussions with our
friends and colleagues Esteban Calzetta, Diego Mazzitelli and
Juan Pablo Paz whose work in the early years contributed toward
the establishment of this field. We acknowledge useful
discussions with Paul Anderson, Larry Ford, Ted Jacobson, Renaud
Parentani and Raphael Sorkin. This work is supported in part by
NSF grant PHY98-00967 and the MyCT and Feder Research Project No.
FPA2001-3598.


\end{document}